\documentclass{emulateapj}

\usepackage{natbib}
\bibpunct{(}{)}{;}{a}{}{,}
\usepackage{graphicx}

\usepackage{amsmath,amssymb}
\usepackage[dvipdfm,breaklinks=true,bookmarks=true,hypertexnames=false,bookmarksnumbered=true,bookmarksopen=false,bookmarks=true,bookmarksnumbered=true]{hyperref}

\newcommand{\msun}{{\rm M}_{\odot}}
\newcommand{\lsun}{{\rm L}_{\odot}}
\newcommand{\kms}{\, {\rm km\, s}^{-1}}
\newcommand{\h}{\,h}
\newcommand{\hmm}{\h^{-1}}
\newcommand{\hmmsun}{\hmm\msun}
\newcommand{\kpc}{\, {\rm kpc}}
\newcommand{\pc}{\, {\rm pc}}
\newcommand{\hmkpc}{\hmm\kpc}
\newcommand{\hmlsun}{\h^{-2}\, \lsun}
\newcommand{\Mpc}{\, {\rm Mpc}}

\newcommand{\der}{{\rm d}}

\newcommand{\mypm}[2]{^{+#1}_{-#2}}
\newcommand{\scrit}{\Sigma_{\rm crit}}

\newcommand{\dos}{D_{\rm os}}
\newcommand{\dol}{D_{\rm ol}}
\newcommand{\dls}{D_{\rm ls}}
\newcommand{\dlsdos}{\dls/\dos}
\newcommand{\eg}{{\it e.g.}~}
\newcommand{\ie}{{\it i.e.}~}
\newcommand{\nbg}{n_{\rm bg}}
\newcommand{\zl}{z_{\rm l}}
\newcommand{\zs}{z_{\rm s}}
\newcommand{\iband}{$I$}
\newcommand{\rein}{R_{\rm Ein}}
\newcommand{\reff}{R_{\rm eff}}
\newcommand{\ecte}{e_{1,{\rm CTE}}}
\newcommand{\dsig}{\Delta \Sigma}
\newcommand{\Sigmabar}{\overline{\Sigma}}

\begin{document}

\title{The Sloan Lens ACS Survey. IV: the mass density profile of
early-type galaxies out to 100 effective radii}

\author{Rapha\"el Gavazzi\altaffilmark{1}}
\author{Tommaso Treu\altaffilmark{1}}
\author{Jason D. Rhodes\altaffilmark{2,3}}
\author{L\'eon V.E. Koopmans\altaffilmark{4}}
\author{Adam S. Bolton\altaffilmark{5}}
\author{Scott Burles\altaffilmark{6}}
\author{Richard J. Massey\altaffilmark{3}}
\author{Leonidas A. Moustakas\altaffilmark{2}}

\altaffiltext{1}{Department of Physics, University of California, Broida Hall, Santa Barbara, CA 93106-9530, USA}
\altaffiltext{2}{Jet Propulsion Laboratory, Caltech, MS 169-327, 4800 Oak Grove Dr., Pasadena, CA 91109, USA}
\altaffiltext{3}{California Institute of Technology, MC 105-24, 1200 East California Boulevard, Pasadena, CA 91125, USA}
\altaffiltext{4}{Kapteyn Astronomical Institute, University of Groningen, PO box 800, 9700 AV Groningen, The Netherlands}
\altaffiltext{5}{Harvard-Smithsonian Center for Astrophysics, 60 Garden St. MS-20, Cambridge, MA 02138, USA}
\altaffiltext{6}{Department of Physics and Kavli Institute for Astrophysics and Space Research, Massachusetts Institute of Technology, 77 Massachusetts Ave., Cambridge, MA 02139, USA}
\email{gavazzi@physics.ucsb.edu}

\shorttitle{Weak lensing around SLACS strong lenses}
\shortauthors{Gavazzi et~al.}

\begin{abstract}
We present a weak gravitational lensing analysis of 22 early-type
(strong) lens galaxies, based on deep {\it Hubble Space Telescope} images
obtained as part of the Sloan Lens ACS Survey. Using advanced
techniques to control systematic uncertainties related to the variable
point spread function and charge transfer efficiency of the Advanced
Camera for Surveys (ACS), we show that weak lensing signal is detected
out to the edge of the Wide Field Camera ($\lesssim 300 \hmkpc$ at the
mean lens redshift $z=0.2$). We analyze blank control fields from the
COSMOS survey in the same manner, inferring that the residual
systematic uncertainty in the tangential shear is less than 0.3\%.  A
joint strong and weak lensing analysis shows that the average {\it
total} mass density profile is consistent with isothermal (i.e. $\rho
\propto r^{-2}$) over two decades in radius (3-300 $\hmkpc$,
approximately 1-100 effective radii). This finding extends by over an
order of magnitude in radius previous results, based on strong lensing
and/or stellar dynamics, that luminous and dark component ``conspire''
to form an isothermal mass distribution.  In order to disentangle the
contributions of luminous and dark matter, we fit a two-component mass
model (de Vaucouleurs + Navarro Frenk \& White) to the weak and strong
lensing constraints. It provides a good fit to the data with only two
free parameters; i) the average stellar mass-to-light ratio
$M_*/L_V=4.48 \pm 0.46 \h\msun/\lsun$ (at $z=0.2$), in agreement with
that expected for an old stellar population; ii) the average virial
mass-to-light ratio $M_{\rm vir}/L_V =
246\mypm{101}{87}\h\msun/\lsun$. Taking into account the scatter in
the mass-luminosity relation, this latter result is in good agreement
with semi-analytical models of massive galaxies formation. The dark
matter fraction inside the sphere of radius the effective radius is
found to be $27\pm4\%$. Our results are consistent with galaxy-galaxy
lensing studies of early-type galaxies that are not strong lenses, in
the region of overlap (30-300 $\hmkpc$). Thus, within the
uncertainties, our results are representative of early-type galaxies
in general.
\end{abstract}


\section{Introduction}\label{sec:intro}

It is now commonly accepted that cold non-baryonic dark matter
dominates the dynamics of the Universe. Whereas the so-called
$\Lambda$CDM (cold dark matter) paradigm has been remarkably
successful at reproducing with high level of precision the properties
of the universe on scales larger than Mpc
\citep[\eg][]{spergel06,tegmark04,seljak05}, the situation at galactic
and subgalactic scales is more uncertain. Dark-matter-only numerical
simulations make very clear predictions. Dark matter halos have a
characteristically ``cuspy'' radial profile
\citep[\eg][]{NFW97,moore98,ghigna98,jing00,stoehr02,navarro04}, are
triaxial \citep{jing02,kazantzidis04c,hayashi06}, and have abundant
substructure \citep[\eg][]{moore99,delucia04,gao04,taylor04}.  From an
observational point of view, substantial effort has been devoted to
comparing those predictions to observations with debated results in the
case \eg of low surface brightness galaxies
\citep{salucci01,deblok03,swaters03,gentile04,simon05} or at galaxy cluster
scales \citep[\eg][]{sand04,gavazzi05,comerford06}.  The main source
of ambiguity in such comparisons is due to the effects of
baryons. Although a minority in terms of total mass, baryons are
dissipative and spatially more concentrated than the dark matter,
playing a critical role at scales below tens of
kiloparsecs. Understanding baryonic physics and the interplay between
dark and luminous matter is necessary to understand how galaxies form
and, ultimately, to test the cosmological model. From an observational
point of view, measuring the relative spatial distribution of stars,
gas, and dark matter is essential to provide clues to help understand
the physical processes and hard numbers to perform quantitative tests
of models.

The dark halos of early-type (i.e. elliptical and lenticular) galaxies
have been particularly hard to detect and study, due to the general
lack of kinematic tracers, such as HI, at large radii. Studies of
local galaxies based on stellar kinematics
\citep{bertin94,gerhard01,mamon05a,mamon05b,cappellari06},
kinematics of planetary nebulae \citep{romanowsky03,arnaboldi04,merrett06}
and temperature profile of X-ray emitting plasma \citep{humphray06} indicate
that at least for the most massive systems dark matter halos are generally
present. The total mass density profile is found to be close to isothermal
($\rho_{\rm tot}\propto r^{-2}$) on scales out to a few effective
radii.

In the distant universe an additional mass tracer is provided by
gravitational lensing. At scales comparable to the effective radius,
strong gravitational lensing makes it possible to detect and study the
mass profile and shape of individual halos \citep{kochanek95} or
statistically of a population of halos \citep{rusin03}. The
combination of strong lensing with stellar kinematics
\citep{treu02,koopmans02,koopmans03,treu04,koopmans06} is particularly
effective, and allows one to decompose the total mass distribution
into a luminous and dark matter with good precision, yielding
information on the internal structure of early-type galaxies all the
way out to the most distant lenses known at $z\sim1$. At larger
scales, the surface mass density is too low to produce multiple
images. However, the weak lensing signal can be detected
statistically by stacking multiple galaxies and measuring the
distortion of the background galaxies. The statistical nature of this
measurement imposes a certain degree of spatial smoothing or binning,
which in turn limits the angular resolution of weak lensing
studies.

In this paper we exploit deep ACS images of 22 gravitational lenses
from the {\it Sloan Lens ACS Survey} \citep[][hereafter papers I, II and III,
respectively]{bolton06,treu06,koopmans06} to perform a joint weak and
strong lensing analysis. This allows us to bridge the gap between the
two regimes and study the mass density profile of early-type galaxies
across the entire range $\sim$1-100 effective radii, disentangling the
luminous and dark components.

From a technical point of view, joint weak and strong lensing analysis
has already been applied in the past to clusters of galaxies and
galaxies in clusters
\citep{natarajan97,geiger98,natarajan02,kneib03,gavazzi03,bradac05a,bradac05b}.
However, there are important differences with respect to galaxy scales. First and
foremost, clusters produce a much stronger weak lensing signal and
therefore it can be detected and studied for individual
systems. Second, since clusters are spatially more extended than
galaxies, relatively large smoothing scales can be adopted to average
the signal over background galaxies. In contrast, the signal of
individual galaxies is too ``weak'' to be detected, so that stacking a
number of lens galaxies is required to reach a sufficient density of
background sources. For this purpose, previous studies have typically
relied on very large sample of galaxies
\citep[][]{brainerd96,griffiths96,wilson01ggl,guzik02,hoekstra04,hoekstra05a,kleinheinrich06} 
or recently in the SDSS survey \citep[][hereafter S04 and M06
respectively]{sheldon04,mandelbaum06}.  As we will show in the rest of
the paper, the high density of useful background galaxies afforded by
deep ACS exposures ($\sim$72 per square arcmin) allows us to achieve a
robust detection of the weak lensing signal with only 22
galaxies, and to study the mass density profile with unprecedented
radial resolution.

The paper is organized as follows. After briefly summarizing the
gravitational lensing formalism and notation in \S~\ref{sec:lensdef},
we discuss the sample, data reduction and analysis, and the main
observational properties of the lens galaxies in
\S~\ref{sec:observ}. \S~\ref{sec:shear-extract} details the shear
measurement, with a particular emphasis on the precision correction for
instrumental systematic effects and on tests of residual systematics
by means of a parallel analysis of blank fields. This section also
presents the mean radial shear profile around SLACS strong lenses and
high resolution two-dimensional mass reconstructions. We combine
strong and weak gravitational lensing constraints in
\S~\ref{sec:modeling} to model the radial profiles lenses and
disentangle the stellar and dark Matter components. We discuss our
results in \S~\ref{sec:discuss} and give a brief summary in
\S~\ref{sec:sum}.

Throughout this paper we assume the concordance cosmological
background with $H_0 = 100\h\,\kms\Mpc^{-1}$, $\Omega_{\rm m}=0.3$ and
$\Omega_\Lambda=0.7$. All magnitudes are expressed in the AB system.

\section{Basic lensing equations}
\label{sec:lensdef}

In this section we briefly summarize the necessary background of
gravitational lensing and especially the weak lensing regime which
concerns the present analysis.
The main purpose of this section is to define notations.
We refer the reader to the reviews of
\citet{mellier99}, \citet{BS01} and \citet{schneider06rev} for more
detailed accounts.

The fundamental quantity for gravitational lensing is the lens
potential $\psi(\vec{\theta})$ at angular position
$\vec{\theta}$ which is related to the surface mass density
$\Sigma(\vec{\theta})$ projected onto the lens plane through
\begin{equation}
\psi(\vec{\theta}) = \frac{4 G}{c^2} \frac{\dol\dls}{\dos} \int \der^2 \theta' \Sigma(\vec{\theta}') \ln\vert \vec{\theta}-\vec{\theta}'\vert\;,
\label{eq:psi}
\end{equation}
where $\dol$, $\dos$ and $\dls$ are angular distances to the lens, to
the source and between the lens and the source respectively. The
deflection angle $\vec{\alpha}=\vec{\nabla}\psi$ relates a point in
the source plane $\vec{\beta}$ to its image(s) in the image plane
$\vec{\theta}$ through the lens equation
$\vec{\beta}=\vec{\theta}-\vec{\alpha}(\vec{\theta})$. The local
relation between $\vec{\beta}$ and $\vec{\theta}$ is the Jacobian
matrix $a_{ij}=\partial \beta_i/\partial \theta_j$
\begin{equation}\label{eq:jacob}
  a_{ij} = \delta_{ij} - \psi_{,ij} = \left(\begin{array}{cc}
      1-\kappa-\gamma_1 & -\gamma_2 \\
      -\gamma_2  & 1-\kappa+\gamma_1
    \end{array}\right)\;.
\end{equation}

The convergence $\kappa(\vec{\theta})=\Sigma(\vec{\theta})/\scrit$ is
directly related to the surface mass density via the critical density
\begin{equation}\label{eq:scrit}
  \scrit = \frac{c^2}{4 \pi G}\frac{\dos}{\dol\dls}\;,
\end{equation}
and satisfies the Poisson equation
\begin{equation}\label{eq:poisson}
\Delta \psi = \psi_{,11}+\psi_{,22} = 2 \kappa \;.
\end{equation}

The 2-component shear is $\gamma=\gamma_1+i \gamma_2 =
\frac{1}{2}(\psi_{,11}-\psi_{,22})+i\psi_{,12}$ in complex
notation. An elliptical object in the image plane is characterized by
its complex ellipticity $e$. In the weak lensing regime, the source
intrinsic ellipticity $e_s$ and $e$ are simply related by $e=e_s+\gamma$.

It is convenient to express the shear into a tangential and a
curl term $\gamma = \gamma_t + i \gamma_\times$ such that $\gamma_t =
-\mathcal{R}\left\{\gamma e^{-2i \varphi}\right\}$ and $\gamma_\times
= -\mathcal{I}\left\{\gamma e^{-2i \varphi}\right\}$ with $\varphi$
the polar angle. For a circularly symmetric lens, $\gamma_\times$ vanishes
whereas $\gamma_t$ at radius $r$ can be written as the difference
between the mean convergence within that radius
$\overline{\kappa}(<r)$ and the local convergence at the same radius
$\kappa(r)$ :
\begin{equation}\label{eq:gammatdef}
  \gamma_t = \overline{\kappa}(<r) - \kappa(r)\,.
\end{equation}
In equations~\ref{eq:psi} to~\ref{eq:gammatdef} we can isolate a
geometric term which linearly scales the lensing quantities $\kappa$,
$\psi$, and $\gamma$ and only depends on the distance ratio
$\dlsdos$. We thus can write $\psi=w(\zl,\zs) \psi_\infty$ (and
analogously for $\kappa$ and $\gamma$) with $w(\zl,\zs)=\dlsdos
\Theta(\zs-\zl)$, where $\Theta(x)$ is the Heaviside step function. If
sources are not confined in a thin plane, we account for the
distribution in redshift by defining an ensemble average distance
factor $\overline{w}(\zl)$ such that:
\begin{equation}\label{eq:zweight}
  \overline{w}(\zl)=\langle w(\zl,\zs)\rangle_{\zs}= \frac{\displaystyle \int_{\zl}^\infty
    \der \zs\, n(\zs) \frac{\dls}{\dos}}
  {\displaystyle\int_0^\infty  \der \zs\, n(\zs)}\,.
\end{equation}

\section{The data}
\label{sec:observ}

\begin{figure*}[htb]
  \centering
  \includegraphics[width=18cm]{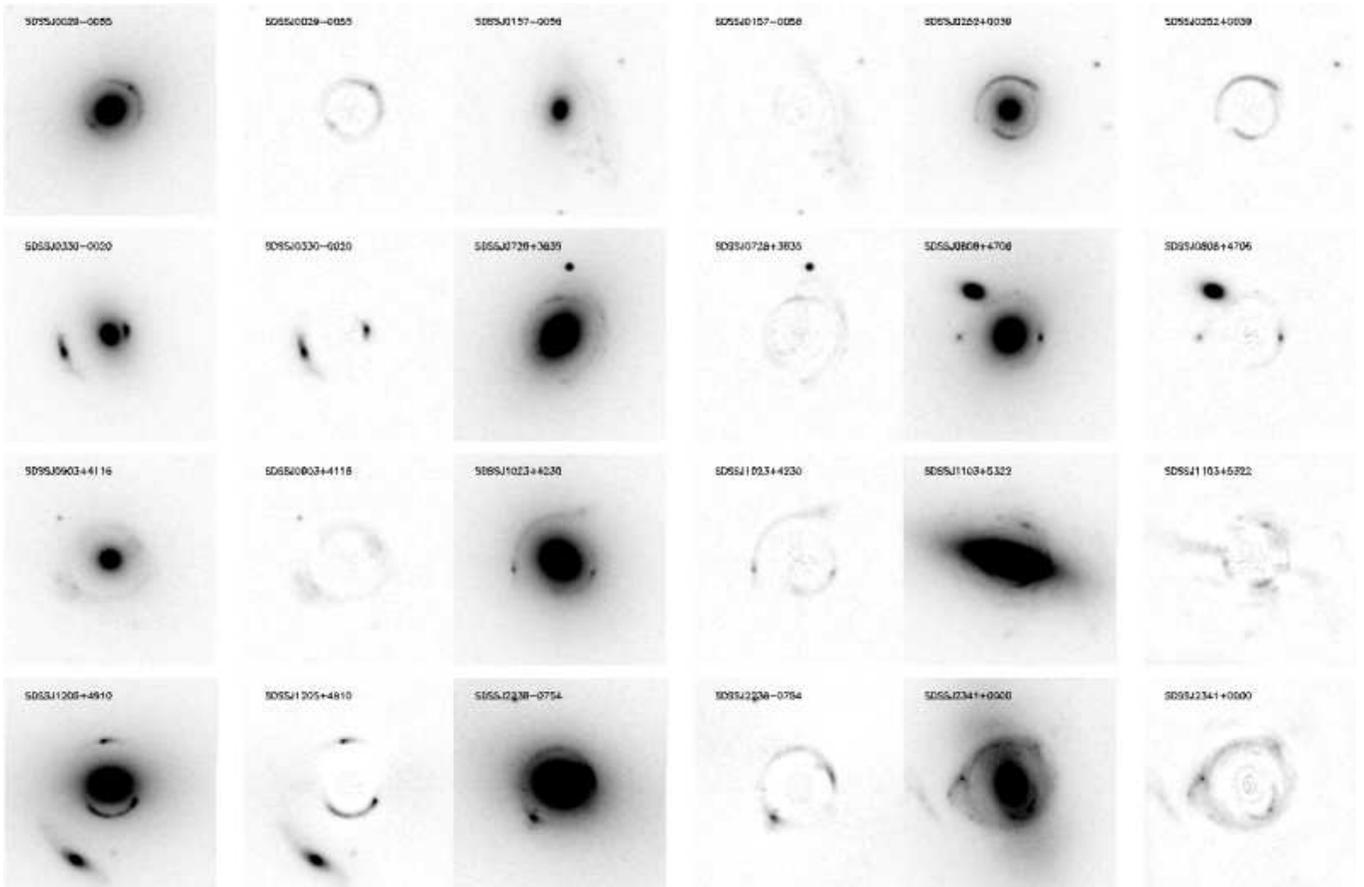}
  \caption{\small F814W gallery of the 12 new SLACS lenses. The
  postage stamps are 7.2 arcseconds on a side. From top to bottom:
  SDSSJ0029-0055, SDSSJ0157-0056, SDSSJ0252+0039, SDSSJ0330-0020,
  SDSSJ0728+3835, SDSSJ0808+4706, SDSSJ0903+4116, SDSSJ1023+4230,
  SDSSJ1103+5322, SDSSJ1205+4910, SDSSJ2238-0754, and
  SDSSJ2341+0000. For each lens we show the original image (left) and
  after subtraction of a model for the lens surface brightness
  (right).}
  \label{fig:newlenses}
\end{figure*}

\subsection{Lens sample}
\label{sec:lenssample}

In this paper we focus on a subsample of 22 lens early-type galaxies
taken from the SLACS Survey (paper I). The subsample is defined as all
the confirmed lenses for which deep, 1-orbit long, ACS image through
filter F814W were available as of the cutoff date for this paper,
2006 October 15.  The parent sample is spectroscopically identified
from the SDSS database and confirmed by ACS imaging, as described in
paper I \citep[see also][]{bolton04}. Ten lenses are in common with
the sample previously analyzed in papers II and III, while the remaining 
12 lenses were not analyzed in papers II and III.
Images of the 12 new lenses are shown in
Figure~\ref{fig:newlenses}. Table \ref{tab:sample} summarizes the most
relevant properties of the 22 lenses. More details on the new lenses
and on the ongoing programs will be presented elsewhere.

The SDSS aperture velocity dispersions are in the range
$196\le\sigma_v\le333 \kms$ and the mean square velocity dispersion is
$\langle\sigma_v^2\rangle^{1/2}\simeq 248\kms$. The lens galaxies have
a mean redshift $\langle \zl=0.22\rangle$. Paper II showed that SLACS
strong lenses fall on the same Fundamental Plane of non-lens
early-type galaxies (see also Bolton et al.\ 2007, in prep). This
demonstrates that -- within our measurement errors -- lensing galaxies
have normal internal dynamical properties at small scales. One of the
goals of this paper is to combine strong with weak lensing to check
whether the outer regions of the SLACS lenses behave in a peculiar way
as compared to non-lens early-type galaxies.

\subsection{HST observations \& data reduction}
\label{ssec:acsdata}

The sample was observed with the Advanced Camera for Surveys on board
HST between November 2005 and October 15 2006, as part of {\it HST} programs
10494 (PI: Koopmans) and 10886 (PI: Bolton). One-orbit exposures were
obtained through filter F814W (hereafter \iband) with the Wide Field
Camera centering the lens on the WFC1 aperture, i.e. in the center of
the second CCD. Hence the observations cover a region as far as
3$\arcmin$ around the lenses.  Four sub-exposures were obtained with a
semi-integer pixel offset ({\tt acs-wfc-dither-box}) to ensure proper
cosmic ray removal and sampling of the point spread function. For the
lenses in program 10494 additional one orbit exposures with ACS
through filter F555W and with the NICMOS NIC2 camera through filter
F160W are also available. In this paper, the additional F555W exposure
is used to check satellite/foreground contamination to the
weak-lensing catalog. A full detailed analysis of the multi-color
images will be presented elsewhere.

Since the goal of this paper is detecting the weak lensing signal
produced by the SLACS strong lenses, we optimize our reduction
according to the prescriptions of \citet[][hereafter R07]{rhodes07}.
For each target, we used {\tt multidrizzle} \citep{koekemoer02} to
combine the four subexposures, using a final pixel size of 0$\farcs$03
and a Gaussian interpolation kernel.

One important difference between this study and that of R07, is that
ours are pointed observations. Thus instrumental effects could play a
different role than for weak-lensing studies of objects at random
positions on the detector, possibly introducing systematic errors. In
order to determine the amplitude of systematic errors in our weak
lensing analysis, we carried out a perfectly analogous analysis of 100
\iband~images of the COSMOS
survey\footnote{\url{http://www.astro.caltech.edu/~cosmos/}} with
identical depth. As detailed below, this allows us to infer the mean
and field-to-field variance of instrumental biases, showing that they
are negligible for our purposes.

\begin{table*}[htbp]
  \caption{Lens sample observed with deep ACS F814W imaging.}\label{tab:sample}
  \begin{tabular}{lccccccccccc}\hline\hline
Name & Prog. ID & Exp. time  & $\zl$  & $\sigma_{\rm ap}$ & $R_{\rm Ein}$ & $R_{\rm eff}$ & $m_I$ &$M_V $ & $\zs$  & $w_{\rm SL}$& $\overline{w}_{\rm WL}$ \\
     &         &   (sec)     &        &  $(\kms)$    &  (arcsec)       & (arcsec)      &  &   $+ 5 \log h$   &     & &     \\\hline
SDSSJ002907.8-005550 & 10886 & 2088 & 0.227 & $228\pm18$ &  0.82 &  1.48 & 17.36 & -21.53 & 0.931  & 0.706  & 0.721 \\
SDSSJ015758.9-005626 & 10886 & 2088 & 0.513 & $295\pm47$ &  0.72 &  0.93 & 18.76 & -22.16 & 0.924  & 0.380  & 0.441 \\
SDSSJ021652.5-081345 & 10494 & 2232 & 0.332 & $333\pm23$ &  1.15 &  2.79 & 16.88 & -22.95 & 0.523  & 0.333  & 0.608 \\
SDSSJ025245.2+003958 & 10886 & 2088 & 0.280 & $164\pm12$ &  0.98 &  1.69 & 17.84 & -21.67 & 0.982  & 0.656  & 0.662 \\
SDSSJ033012.1-002052 & 10886 & 2088 & 0.351 & $212\pm21$ &  1.06 &  1.17 & 18.20 & -21.86 & 1.107  & 0.613  & 0.589 \\
SDSSJ072805.0+383526 & 10886 & 2116 & 0.206 & $214\pm11$ &  1.25 &  1.33 & 16.95 & -21.80 & 0.688  & 0.660  & 0.745 \\
SDSSJ080858.8+470639 & 10886 & 2140 & 0.219 & $236\pm11$ &  1.23 &  1.65 & 17.10 & -21.77 & 1.025  & 0.735  & 0.730 \\
SDSSJ090315.2+411609 & 10886 & 2128 & 0.430 & $223\pm27$ &  1.13 &  1.28 & 18.19 & -22.25 & 1.065  & 0.521  & 0.512 \\
SDSSJ091205.3+002901 & 10494 & 1668 & 0.164 & $326\pm12$ &  1.61 &  5.50 & 15.20 & -22.95 & 0.324  & 0.472  & 0.794 \\
SDSSJ095944.1+041017 & 10494 & 2224 & 0.126 & $197\pm13$ &  1.00 &  1.99 & 16.61 & -20.94 & 0.535  & 0.738  & 0.840 \\
SDSSJ102332.3+423002 & 10886 & 2128 & 0.191 & $242\pm15$ &  1.30 &  1.40 & 19.93 & -21.56 & 0.696  & 0.686  & 0.762 \\
SDSSJ110308.2+532228 & 10886 & 2156 & 0.158 & $196\pm12$ &  0.84 &  3.22 & 16.02 & -22.02 & 0.735  & 0.749  & 0.801 \\
SDSSJ120540.4+491029 & 10494 & 2388 & 0.215 & $280\pm13$ &  1.04 &  1.92 & 16.76 & -22.00 & 0.481  & 0.521  & 0.735 \\
SDSSJ125028.3+052349 & 10494 & 2232 & 0.232 & $252\pm14$ &  1.15 &  1.64 & 16.78 & -22.17 & 0.795  & 0.662  & 0.716 \\
SDSSJ140228.1+632133 & 10494 & 2520 & 0.205 & $267\pm17$ &  1.39 &  2.29 & 16.44 & -22.20 & 0.481  & 0.543  & 0.747 \\
SDSSJ142015.9+601915 & 10494 & 2520 & 0.063 & $205\pm04$ &  1.04 &  2.49 & 14.93 & -21.04 & 0.535  & 0.867  & 0.919 \\
SDSSJ162746.5-005358 & 10494 & 2224 & 0.208 & $290\pm14$ &  1.21 &  2.47 & 16.79 & -22.06 & 0.524  & 0.570  & 0.743 \\
SDSSJ163028.2+452036 & 10494 & 2388 & 0.248 & $276\pm16$ &  1.81 &  2.01 & 16.76 & -22.31 & 0.793  & 0.639  & 0.698 \\
SDSSJ223840.2-075456 & 10494 & 2232 & 0.137 & $198\pm11$ &  1.20 &  2.33 & 16.20 & -21.58 & 0.713  & 0.776  & 0.827 \\
SDSSJ230053.2+002238 & 10494 & 2224 & 0.228 & $279\pm17$ &  1.25 &  2.22 & 16.91 & -22.06 & 0.463  & 0.476  & 0.719 \\
SDSSJ230321.7+142218 & 10494 & 2240 & 0.155 & $255\pm16$ &  1.64 &  3.73 & 15.97 & -22.40 & 0.517  & 0.670  & 0.805 \\
SDSSJ234111.6+000019 & 10886 & 2088 & 0.186 & $207\pm13$ &  1.28 &  3.20 & 16.30 & -22.14 & 0.807  & 0.729  & 0.768 \\ \hline
  \end{tabular}\\

Apparent I band magnitudes are not corrected for Galactic extinction.
Absolute magnitudes are K-corrected, extinction corrected, and
corrected to the sample mean redshift $z=0.2$ for luminosity evolution
using $\log L_V,0.2=\log L_V,z+0.40*(z-0.2)$. Combining measurement
errors and uncertainties in various photometric corrections yields a
typical error in apparent (resp. absolute) magnitudes $\pm0.02$
(resp. $\pm0.04$) mag. Relative uncertainties in $\rein$ are about
$5\%$ and $\sim 10\%$ for $\reff$. Since systems are elliptical,
both $\rein$ and $\reff$ are expressed relative to the geometric
mean intermediate radius. $w_{\rm SL}$ is the lensing
distance ratio for the strong lensing event source redshift whereas
$\overline{w}_{\rm WL}$ is the same distance ratio averaged over the
redshift distribution of background sources used for weak lensing.
\end{table*}

\subsection{Surface photometry and lens models}

Surface photometry of the lens galaxies was obtained by fitting de
Vaucouleurs profiles after carefully masking the lensed structures
(rings) and any neighboring bright satellites. The two-dimensional
parametric fit was carried out using {\tt galfit} \citep{peng02}. We
checked that our results are consistent with those of paper II for the
10 lenses previously observed with shallower HST snapshot imaging
\citep[see corrected table 2 of paper II in][]{treu06erratum}.

We determined absolute V band magnitudes of the lenses taking into
account filter transformations and galactic extinction according to
the \citet{schlegel98} dust maps. Furthermore, in order to homogenize
the sample, we passively evolved all V band magnitudes to a fiducial
redshift $z=0.2$, using the relation (\citet{treu01} and paper II):
\begin{equation}\label{eq:dlogmlv}
  \frac{\der \log \frac{M_*}{L_V}  }{\der z} \simeq -0.40 \pm 0.05\;,
\end{equation}
which is well suited for the massive early-type galaxies we are
considering here. We note that the correction is of order a few
hundredths dex, and adopting a different passive evolution would not
alter our results in any significant way. Thus the V-band luminosities
listed in Table~1 are $z=0.2$ V band luminosities and can be considered
as fair proxies for the lens stellar mass up to an average stellar
mass-to-light ratio $\Upsilon_V\equiv M_*/L_V$.

We measured Einstein radii in full analogy to paper III, that is, we
parameterized the lens potential with a Singular Isothermal Ellipse
profile and reconstructed the unlensed source surface brightness
non-parametrically to match the observed Einstein ring features.
Typical uncertainties on the recovered values of $\rein$ are $\sim
0.05$ with small variation between lenses. Again, we checked
that the present modeling provides consistent results with respect to
those in paper III. A more detailed description of strong lensing
modeling of the 12 new lenses will be given in forthcoming papers.

\section{Shear analysis}\label{sec:shear-extract}

\subsection{Background sources selection}
\label{sec:shear:gen}

The detection of background sources was done with {\tt
imcat}\footnote{\url{http://www.ifa.hawaii.edu/~kaiser/imcat/}} and
cross-correlated with the {\tt SExtractor} \citep{bertin96} source
catalog to remove spurious detections. To limit screening by the
foreground main lens we subtracted its surface brightness profile
before source detection with {\tt SExtractor} and {\tt imcat}.  After
identifying stars in the magnitude size diagram in a standard manner,
we applied the following cuts to select objects for which shapes could
be reliably measured. First, we restricted the analysis to galaxies
brighter than $I<26$ although the galaxy sample is complete down to
$I\sim27.5$, based on the number counts. This removes faint and small
objects with poorly known redshift distribution. Second, we applied a
bright $I\ge20$ cut to the source sample, to minimize foreground
contamination. Third, we discarded objects with an half-light radius
$r_h\le0.09\arcsec$ (for comparison the \iband~PSF has
$r_h\sim0.06\arcsec$).  Fourth, pairs of galaxies with small angular
separation ($\le 0.5\arcsec$) were discarded since their shape cannot
be reliably measured.  After these cuts, we achieve a number density
of useful background sources $\nbg=72\: {\rm arcmin}^{-2}$.

The redshift distribution of sources is taken from the recently
published COSMOS sample of faint galaxies detected in the ACS/F814W
band \citep{leauthaud07}. Their analysis exploits photometric
redshifts, to rerive redshift sources distribution down to magnitude
$I\simeq 26$ for a sizeable sample selected at HST resolution.
The redshift distribution of
sources having $\iband\le 26$ is well represented by the following
expression
\begin{equation}\label{eq:zsdistrib}
  \frac{\der n(\zs)}{\der \zs} = \frac{1}{z_0 \Gamma(a)} e^{- \zs/z_0} (\zs/z_0)^{a-1}\;, 
\end{equation}
with $z_0=0.345$ and $a=3.89$. For this particular redshift
distribution, values of $\overline{w}_{\rm WL}=\langle \dlsdos
\rangle_{z_s}$ are reported in Table \ref{tab:sample}.  This redshift 
distribution represent a clear improvement of our analysis over
previous estimates based on ground based surveys, as the redshift
distribution of faint sources depends not only on magnitude but also
on object size.  The relatively low redshifts of SLACS lenses and the
rapid saturation of $w(\zl,\zs)$ with increasing source redshift helps
reduce the sensitivity of our results to residuals errors on
photometric redshifts. Taking into account current errors on $\der
n(\zs)/\der \zs$ reported by \citet{leauthaud07}, the overall
calibration for our sample is accurate to $\sim7\%$. In the rest of
the paper we will show that this uncertainty is negligible for our
purposes.

A potential additional concern is residual contamination by satellites
galaxies that are spatially correlated with the main lens galaxy and
thus could dilute the weak lensing signal. Furthermore, we expect the
number of satellites to depend on the distance from the lens center,
and this could potentially affect our inferred shear profile.

As a first check, we applied a color cut to the background catalog of
the 10 SLACS fields for which F555W filter imaging is available. We
measured the shear signal for galaxies redder than the lenses (\ie
$F555W-F814W \ge 1.5$), expected to be at higher redshift \citep[see
\eg][for similar color selections]{broadhurst05b,limousin06}. The
signal-to-noise on the recovered shear profile for this tiny subsample
of sources turned out to be too small for this test to be
conclusive. This test will be more powerful when the full multi-color
dataset will be available at the end of the survey.

Therefore we turned to comparisons with the weak lensing SDSS analysis
of S04 who found that about 10\% of $r<22$ sources are correlated to
the lens at scales $\sim 30\kpc$. Our ACS catalogs are 4 magnitudes
deeper than SDSS catalogs. The number density of background sources is
much higher $N_{\rm bg}(I<26)/N_{\rm bg}(I<22)\sim46$ but the number
of satellites should also increase. Assuming a typical luminosity function
with slope $\alpha=-1$, we can
extrapolate our counts and predict that $N_{\rm sat}(I<26) / N_{\rm
sat}(I<22) \lesssim 2$. Therefore, at $\sim 30 \kpc$ from the lens
center, the contamination must be at most $N_{\rm sat}/N_{\rm bg}
\simeq 10 \times 2 / 46 \sim 0.5\%$. Similarly, at smaller scale
$r\sim3\kpc$, we can extrapolate S04 results to predict that the
contamination ratio increases by at most a factor 10, yielding $N_{\rm
sat}/N_{\rm bg} \simeq 5\%$. This ratio, as we will see below, is much
smaller than present error bars ($\gtrsim 30\%$ per bin) so we conclude
that contamination by satellites cannot be a significant source of
error. This finding is supported by the excellent agreement between
strong and weak lensing measurements at small scales (see below).

\subsection{Instrumental distortions}
\label{sec:shear:distro}

Before using the shape of background galaxies as a tracer of the shear
field, we need to correct several instrumental effects. Because every
lens galaxy is approximately\footnote{Typically within a few pixels
due to absolute pointing uncertainties. The stack is of course aligned
on the measured center of each galaxy.} at the same location in the
detector frame (in the middle of CCD2) we need to carefully assess and
correct any instrumental source of systematic polarization of galaxies
which may bias the measured shear profile. To correct for the smearing
of galaxy shape by the Point Spread Function (PSF), we use the well
known KSB method \citep{KSB95} which has proved being a reliable
method down to cosmic shear requirements \citep{heymans06step}.  The
implementation we are using is similar to that of
\citet{gavazzi06clus} but is tuned for the specific space-based
conditions \citep[see \eg][]{hoekstra98} by adaptively matching
the radial size of the weight function applied to stars to that of the
galaxies that are being PSF-corrected. Some important changes
inspired by R07 are detailed in the following \citep[see
also][for further discussion of the techniques required to extract
weak lensing signal from ACS images]{schrabback06}.

\subsubsection{Focus \& Point Spread Function smearing}
\label{sec:shear:psf}

The shape of galaxies must be corrected for the smearing by the Point
Spread Function of the ACS camera which circularizes objects and/or
imprints systematic distortion patterns. The PSF from space-based
images is expected to be more stable as compared to ground-based data
which suffer from time varying atmospheric seeing conditions. However,
R07 showed that the ACS PSF varies dramatically as a function of time,
essentially because of focus oscillations due to thermal
breathing. The peculiar off-axis position of ACS enhances any focus
variability and PSF anisotropy is difficult to control. Unfortunately,
we cannot map PSF variations across the field from the data
themselves, since not enough stars are observed in each exposure.
Therefore, following R07, we compare the few available stars to mock
PSFs built with {\tt TinyTim} \citep{Krist04} and modified as
described in R07 as a function of focus and detetmine the best fitting
focus. The distribution of offsets $\delta_{\rm focus}$ with respect
to the nominal ACS focal plane is well consistent with R07 \ie
$\delta_{\rm focus} \sim -3.4 \pm 0.8\,\micron$. As a consistency
check we apply the same procedure to the blank fields from COSMOS.

\begin{figure}[tbh]
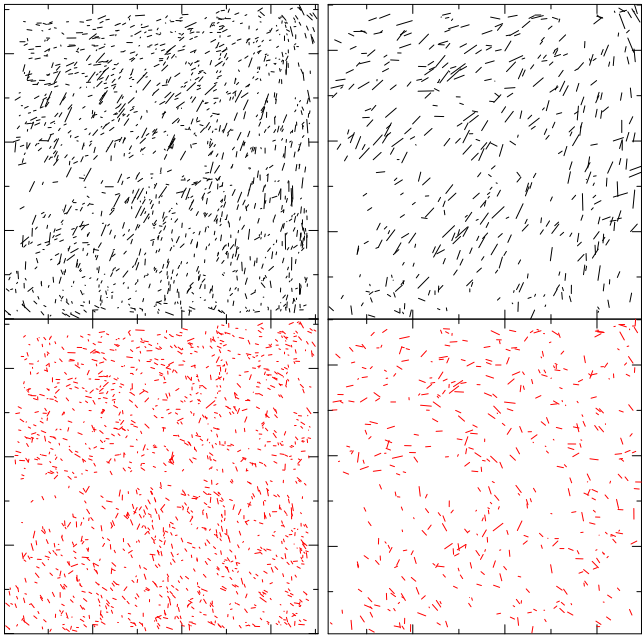

  \centering
  \includegraphics[width=4.2cm,height=8.4cm]{psf_anis_cosmos.eps}
  \includegraphics[width=4.2cm,height=8.4cm]{psf_anis_slacs.eps}

  \caption{\small Ellipticity patterns measured onto stars in {\it
  all} the COSMOS (left) and SLACS (right) fields. Upper {\it (black)}
  panels show stellar ellipticities before applying PSF correction
  scheme whereas the lower {\it (red)} panels show that the main
    distortion patterns are significantly suppressed after correction.
  Each panel corresponds to the full ACS field of view
  (\ie $200\arcsec$ on a side)}
  \label{fig:psf-check}
\end{figure}

Fig. \ref{fig:psf-check} shows the ellipticity of stars before and
after our PSF correction scheme for the 100 COSMOS fields and our 22
SLACS fields. Averaged over the COSMOS fields we measured a mean
complex ellipticity $\langle e_{*,{\rm uncor}} \rangle = (-0.0037 +
0.0072 i ) \pm 0.0004$ before PSF anisotropy correction and $\langle
e_{*,{\rm cor}} \rangle = (0.0000 - 0.0043 i ) \pm 0.0003$ after
correction. Similarly, in SLACS fields we obtained $\langle e_{*,{\rm
uncor}} \rangle = (-0.0025 + 0.0096 i ) \pm 0.0006$ and $\langle
e_{*,{\rm cor}} \rangle = ( 0.0011 - 0.0023 i ) \pm 0.0005$. In both
datasets the scatter of corrected stellar ellipticities about this
mean is isotropic and $\sigma_{e_{*,{\rm cor}}}\simeq 0.0087$. We
conclude that the correction reduces the {\it mean} PSF anisotropy by
a factor $2-4$ although some residuals are still present at the
$\sim 0.003$ level. We will show in \S~\ref{sec:shear:check} that the
residual uncertainties are negligible for our purpose.

In addition to anisotropic distortions, the convolution with the PSF
also produces isotropic smearing making objects appear rounder. This
effect is much smaller on HST images than from the ground but it must
be taken into account for small objects with a typical size comparable
to that of the PSF. Our initial size cut $r_h\ge0.09\arcsec$
guarantees that such isotropic smearing is kept at a low level
and the KSB method can perform an accurate correction
\citep{KSB95,hoekstra98}.  Our implementation of KSB builds on the
proposed improvements suggested by the STEP1 and STEP2 results
\citep{heymans06step,massey07step}.  These papers indicate that, in general, 
KSB methods can achieve $\sim$10\% relative shear calibration biases
or smaller. Since this is smaller than our statistical errors, it is
sufficient to adopt for the present paper a conservative 10\%
systematic uncertainty in our shear calibration ($m$ STEP
parameter). In a future paper we plan to take advantage of the future
{\it spaceSTEP} simulation set to get a more accurate estimate of the 
uncertainty on the shear
calibration. This will be necessary given the gain in sensitivity
expected when the deep SLACS follow-up will be complete. In addition,
we demonstrate in the next section that no significant residual
additive term ($c$ STEP parameter) is observed in either the SLACS
data or the COSMOS control fields.

\subsubsection{Charge Transfer Efficiency}
\label{sec:shear:cte}

Another source of systematic distortion is the degradation of Charge
Transfer Efficiency (CTE) on ACS CCDs. Charges are delayed by defects
in the readout direction (\ie~the $y$ axis, charges going from the gap
between the CCD chips toward the field boundaries), imprinting a tail
of electrons behind each object which modifies its shape, thus
producing a spurious negative $e_1$ component. Since the strength of
CTE-induced distortions depends on the signal-to-noise (snr) of the
objects (the fainter the source the stronger the effect), we cannot
measure distortions from stars and correct faint galaxies
accordingly. This effect must be quantified and corrected with
galaxies themselves but one must be able to distinguish between the
physical signal and the CTE distortions. To this aim we use the blank
COSMOS fields to make sure that our CTE correction scheme will
efficiently remove CTE distortions while leaving the actual shear
signal unchanged. In practice, we use
the empirical recipe proposed by R07 in which an $\ecte$ component,
function of $y$ pixel coordinate, snr, observation time (since CCD
degradation increases with time) is subtracted for each object. Here
we adapt the expression from Eq. (10) of R07
\begin{equation}\label{eq:CTE}
\begin{split}
  \ecte = -3.6\times 10^{-4} \left( \frac{1}{2}- \left\vert \frac{1}{2}- y' \right\vert \right)\,{\rm snr}^{-0.9}\, \times \, \\
  \left( {\rm MJD} -52333 \right) \left(\frac{r_h}{0.18 \arcsec}\right)^{-0.1}
\end{split}
\end{equation}
to the {\tt imcat} definition of snr whereas R07 use {\tt
SExtractor}. Note that the small dependence of $\ecte$ on size is
somewhat degenerate with the way snr is defined and may not be always
necessary (like in R07). Note also that snr is calculated by {\tt
imcat} without taking into account noise correlation caused by
multidrizzling on oversampled pixels ($0.03\arcsec$ pixel size instead
of the native $0.05\arcsec$ value). Therefore this expression may not
be directly applicable with different {\tt multidrizzle} settings. MJD
is the Modified Julian Date of observation, $r_h$ is the half-light
radius and $y'$ is the normalized $y$ pixel coordinate\footnote{$y'=0$
(resp. 1) at the bottom (resp. top) edge of the ACS field of view.}.

\begin{figure*}[htb]
  \centering
  \includegraphics[width=8.5cm,height=10.cm]{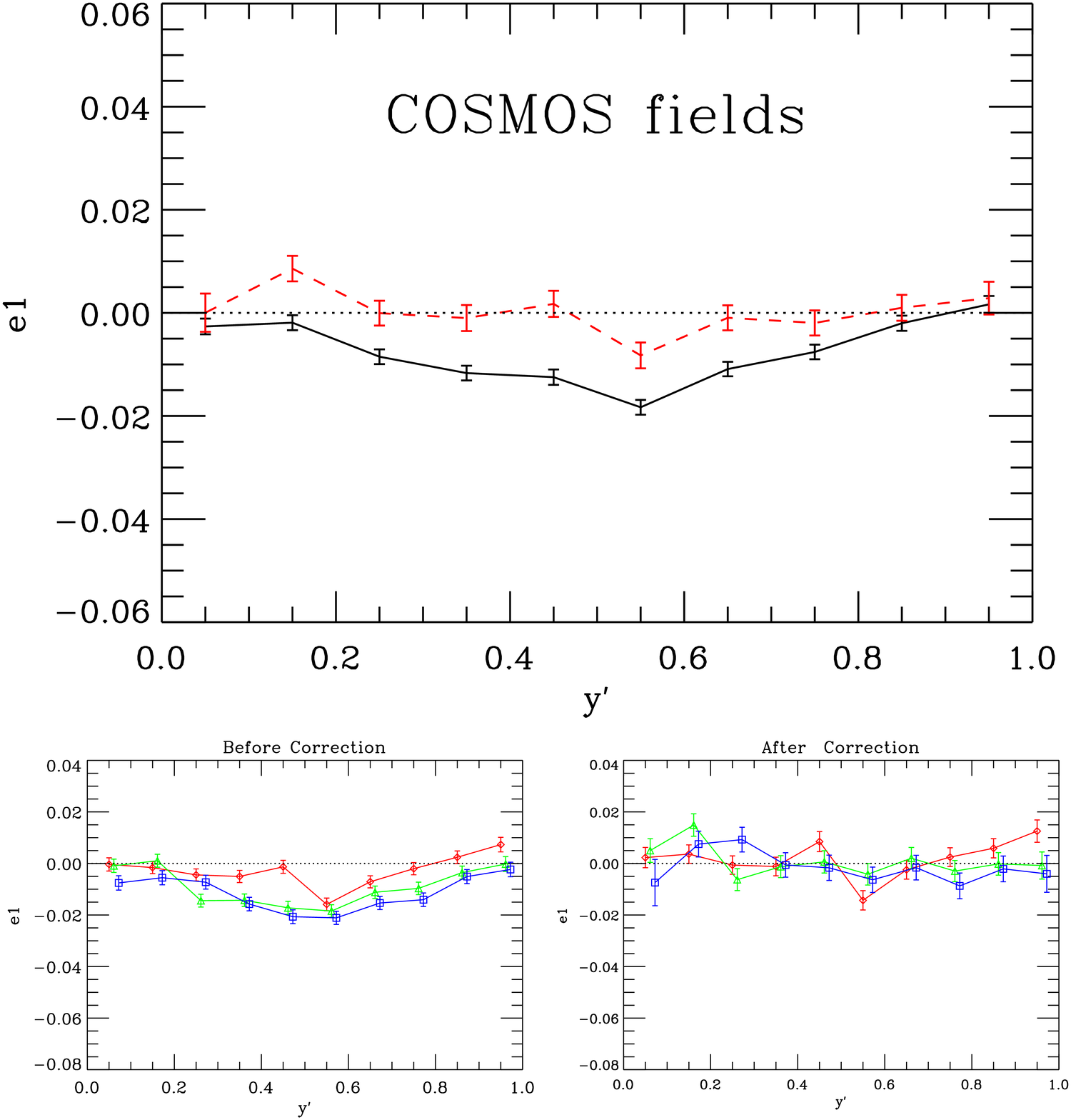}
  \includegraphics[width=8.5cm,height=10.cm]{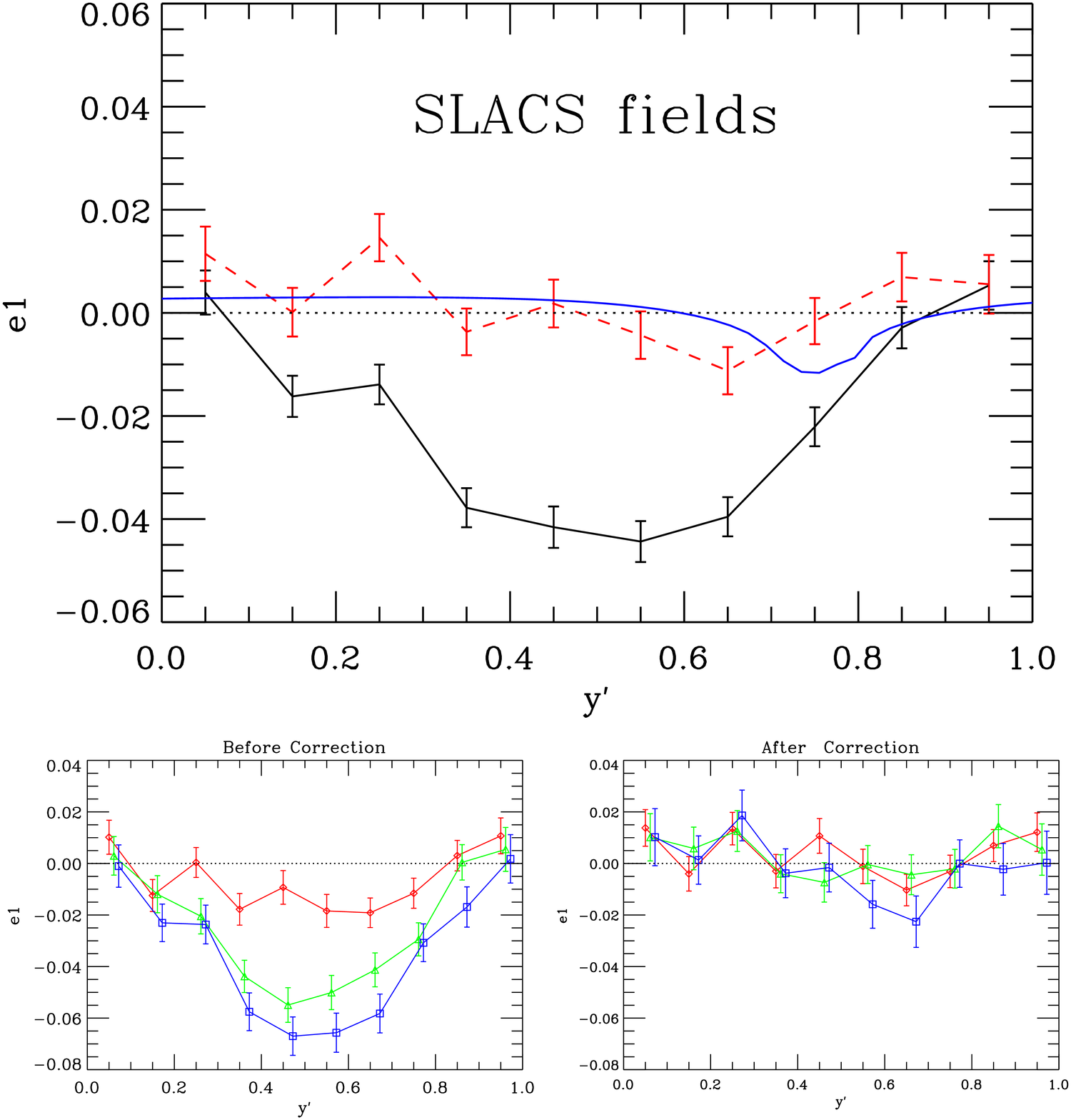}

  \caption{\small Vertical spurious stretching (negative $e_1$
  ellipticity component) as a function of vertical pixel coordinate
  due to CTE degradation. Measurements for COSMOS and SLACS fields
  are shown in the left and right panels respectively.
  \ \ \ \ 
  {\it Upper panels:} average stretching before (solid black) and
  after (dashed red) applying our empirical correction
  term. Distortions are maximum far from the readout (gap between CCDs
  at $y'\sim0.5$).  They are stronger in SLACS images because the
  number of defects in CCDs increases with time.  The corrected
  residual signal should not perfectly ``flat'' in SLACS fields
  because of the presence of the true weak lensing signal we aim at
  detecting. The amplitude of the mean $e_1$ shear signal expected
  from an isothermal sphere with Einstein Radius $\rein=1.2\arcsec$ is
  shown for comparison (solid blue).
  \ \ \ \ The {\it lower left and lower right panels} show that
  distortions are stronger for fainter objects as illustrated by
  splitting galaxies into three magnitude bins ({\it red, green, blue}
  curves for bright, intermediate and faint objects). Comparing $e_1$
  ellipticities before ({\it left}) and after ({\it right}) applying
  our correction scheme, we see that equation \eqref{eq:CTE} corrects
  this spurious signal at all magnitudes.}  \label{fig:cte-check}
\end{figure*}

Fig. \ref{fig:cte-check} illustrates the effect of CTE on $e_1$
ellipticity components for COSMOS and SLACS fields (respectively left
and right panels). We show the $e_1$ component of galaxy ellipticity
as a function of the $y'$ frame coordinate for sources brighter than
$\iband=27$ (\ie well beyond our magnitude cut for selecting suitable
sources).  The upper panels show the mean $e_1$ before and after
CTE(+PSF) correction. For COSMOS and SLACS we see a
similar tendency for vertical stretching of galaxies in the middle of
the frame. The empirical CTE distortion fitting formula \eqref{eq:CTE}
provides a good correction. Although statistical errors are larger in
the SLACS images (5 times smaller sample) we see a modulation of the
corrected $e_1$ component as a function of $y'$ that is not observed
in the corrected COSMOS images. As we shall see below this is the
signature of the signal we are interested in and it should not be
erased by the CTE correction scheme (in order to compare this residual
with the expected shear, we overlay the $\langle e_1(y')\rangle$ shear
signal from an isothermal sphere with Einstein Radius
$\rein=1.2\arcsec$). The lower panels (left and right) show how CTE
distortions depend on signal to noise ratio.  We have split the galaxy
sample into three magnitude quantiles, the bright objects being less
distorted. This is well accounted for by the snr (and size) dependence
in Eq. \eqref{eq:CTE}. The overall amplitude of CTE distortion is
approximately double for SLACS images because of increasing CCD degradation
with time. This is also well captured by Eq. \eqref{eq:CTE}. The
median observation date of the 100 COSMOS exposures we are considering
is $\overline{MJD}_{\rm cosmos}=53141$ and for SLACS it is
$\overline{MJD}_{\rm slacs}=53972$.

\subsubsection{Checks on residuals}
\label{sec:shear:check}

\begin{figure*}[htb]
  \centering
  \includegraphics[width=16cm,height=13cm]{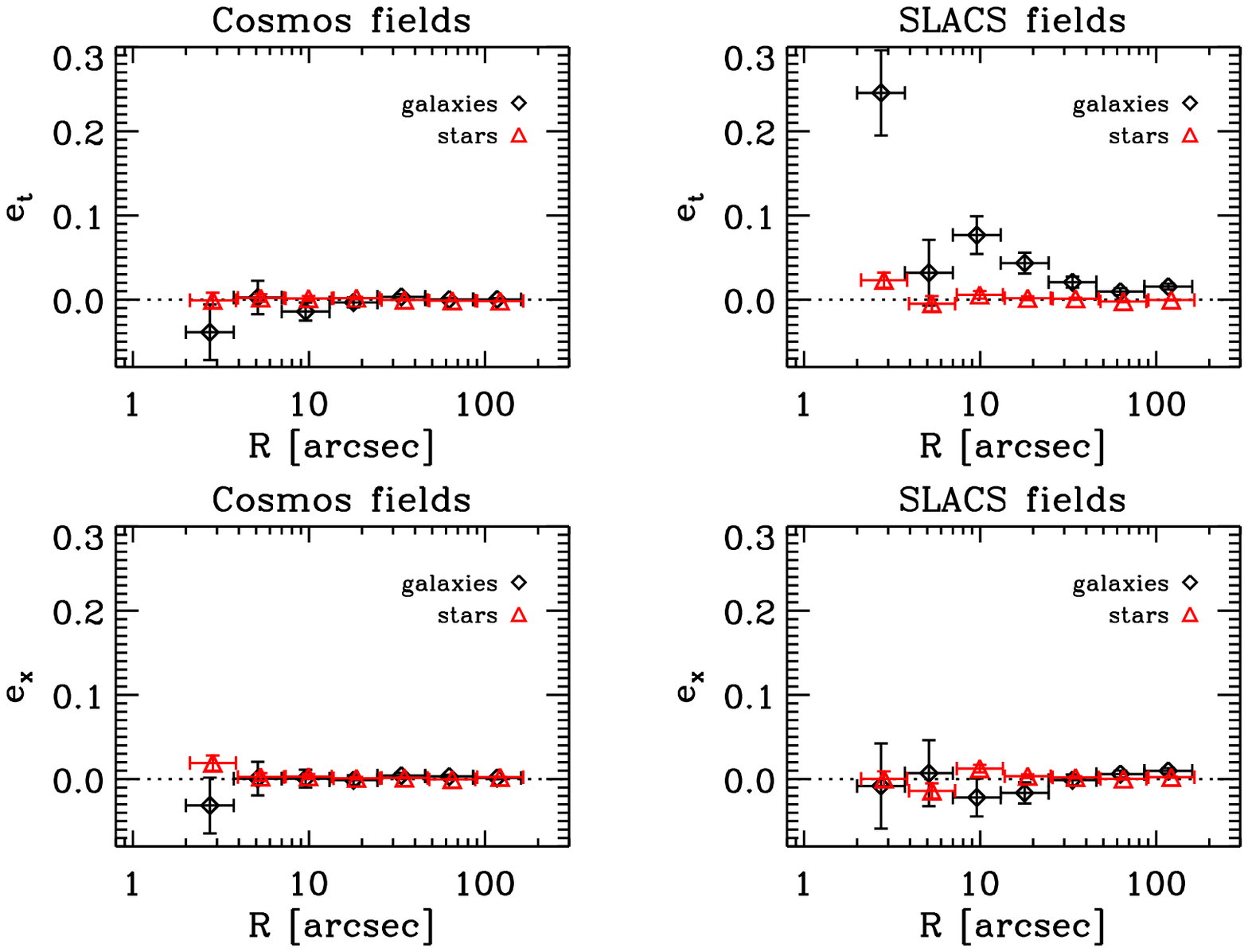}

  \caption{\small Radial ellipticity profile in COSMOS {\it left} and
  SLACS {\it right} fields. Upper panels show the tangential component
  $e_t$ profile and lower panels show the curl component
  $e_\times$. The signal is measured on both distortion-corrected
  stars {\it (red triangles)} and galaxies {\it (black squares)}. For
  plotting convenience the innermost data point in the upper right
  panel ($e_t$ in SLACS) falls outside the window.}
  \label{fig:rawshearprof}
\end{figure*}

To test the quality of the instrumental systematics correction, we plot in
Fig. \ref{fig:rawshearprof} the radial profile of both the tangential
$e_t$ and curl $e_\times$ components of the complex ellipticity in
COSMOS and SLACS fields (left and right respectively). The center is
set on the lens for SLACS images and at the same location in the
detector frame in COSMOS. If we first focus on these latter images, we
see no statistically significant residual $e_t$ nor $e_\times$
component, thus showing that we are free from PSF (seen in stars and
galaxies) or CTE (seen in galaxies only) systematics. As well around
SLACS lenses, stars do not carry any significant residual $e_t$ or
$e_\times$ signal. Therefore we can safely assume that our systematics
correction scheme is accurate enough for the present
analysis\footnote{At the end of the SLACS deep survey, we expect
$\sim$80-100 lenses. Since systematics are already below statistical
errors in COSMOS (100 fields), our treatment is satisfactory for the
final sample.}. Galaxies in SLACS fields clearly carry a strong $e_t$
signal (note the first data point well outside the plotting window)
whereas no statistically significant $e_\times$ component is observed
as expected for a gravitational lensing origin of this shear signal.

\subsection{Other sources of error}
\label{sec:shear:other}

A final additional potential source of systematic uncertainy is the
effect of the lens galaxy surface brightness on the ellipticity of
background sources at small projected radii. To estimate this effect,
we detected and measured object shapes before and after subtraction of
the lens surface brightness profile using {\tt galfit}
and b-spline techniques developed for the strong lensing
analysis (paper I). These two method yield similar results for
the purpose of weak lensing. It turns out that the lens-subtraction
process changes measured ellipticities by at most 5\% in an incoherent
way. Therefore we do not consider further the effect of lens surface
brightness as a relevant potential source of systematic.

\begin{figure}[htb]
  \centering
  \includegraphics[width=11.cm]{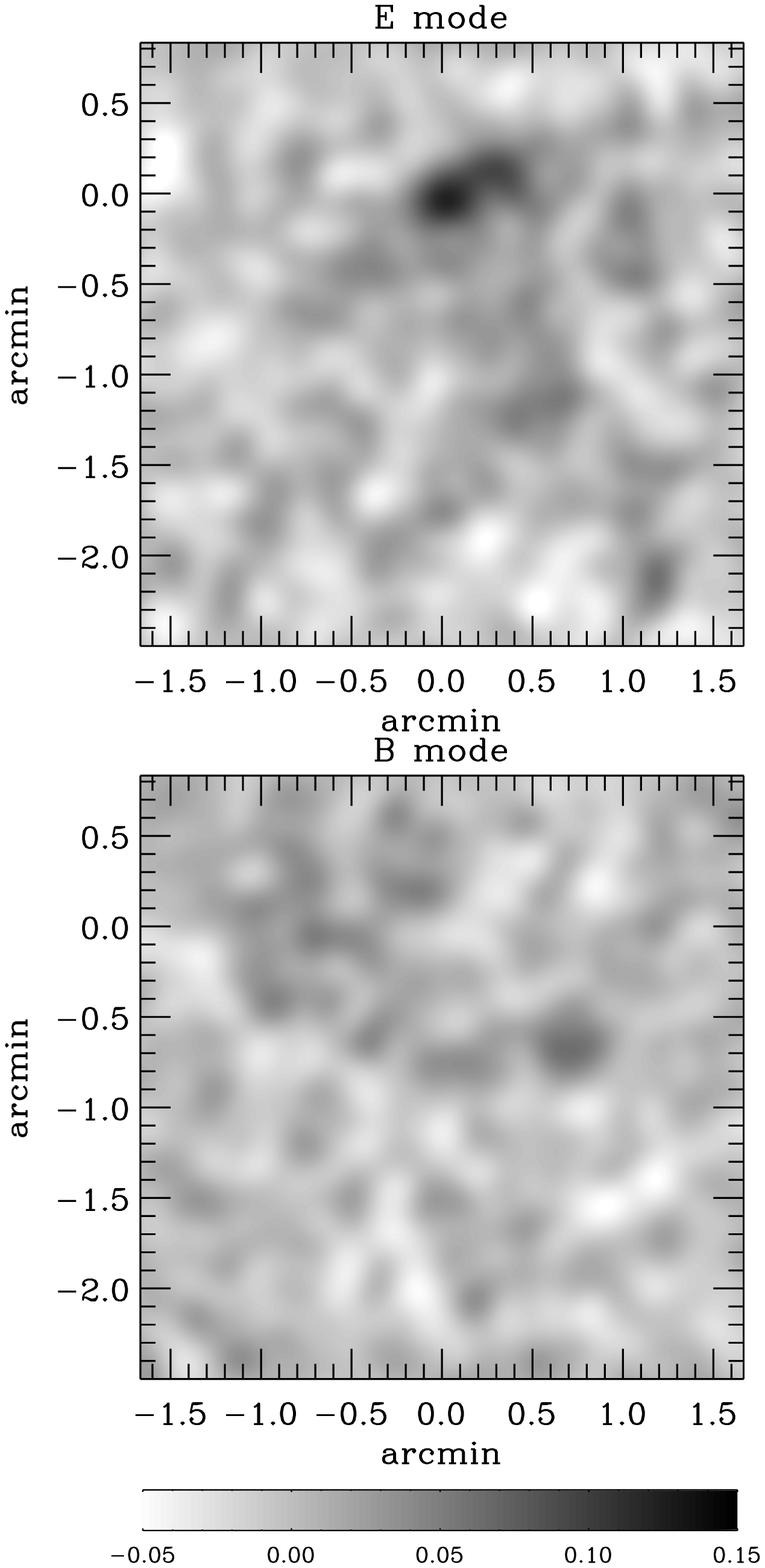}
  \caption{\small Mass reconstruction around stacked SLACS lenses in
  the same detector frame. The {\it upper panel} shows the convergence
  map (E-mode) whereas the {\it lower panel} shows the recovered
  convergence map after rotating galaxies by $45^\circ$ (B-mode) which
  is consistent with a pure noise realization.}
  \label{fig:massmap}
\end{figure}

\subsection{Two-dimensional mass reconstruction and shear profile}
\label{sec:prof-maps}

In the upper panel of Fig. \ref{fig:massmap} a mass reconstruction
(convergence $\kappa$ map) around the stacked lenses using the
\citet{kaiser93} method is shown. There is only one significant
convergence peak at the position of the main lens. Note that the
Gaussian smoothing scale of the convergence maps is 8 arcsec. This
extraordinary high spatial resolution is made possible by the high
density of background sources. The lower panel shows the imaginary
part of the reconstructed mass map (obtained after rotating background
galaxies by $45^\circ$). This is consistent with a pure noise map and
illustrates the amplitude of the noise.

We now analyze the radial shear profile achieved by stacking the lens
galaxies. Since the images are taken at a random orientation with
respect to the lens major axis we can safely assume circular symmetry
in the analysis. We convert the shear $\gamma$ into the physical
quantity $\dsig(R) = \scrit \gamma (R) = \Sigmabar(<R)-\Sigma(R)$.
For a given lens redshift $\zl$ we also define the average critical density
$\scrit'=\frac{c^2}{4 \pi G} \frac{1}{\dol \overline{w}(\zl)}$. An
estimator for $\dsig$ at a given radius is
\begin{equation}\label{eq:dsigdef}
  \dsig = \frac{\sum_{j=1}^{N_{\rm lens}}{\scrit'}_j^{-1} \sum_{i=1}^{N_{s,j}} e_{t,i} \sigma_{e,i}^{-2} }{\sum_{j=1}^{N_{\rm lens}}{\scrit'}_j^{-2} \sum_{i=1}^{N_{s,j}} \sigma_{e,i}^{-2} }
\end{equation}
where $N_{s,j}$ is the number of sources in the radial bin around lens
$j$ and $\sigma_{e,i}$ is the uncertainty assigned to the tangential
ellipticity estimate $e_{t,i}$ \citep[see][for details on this
weighting scheme]{gavazzi06clus}. With this definition, $\dsig$ is
directly comparable to other SDSS weak galaxy-galaxy lensing analyses
(\eg S04, M06).

\begin{figure}[htb]
  \centering
  \includegraphics[width=8.6cm,height=9.0cm]{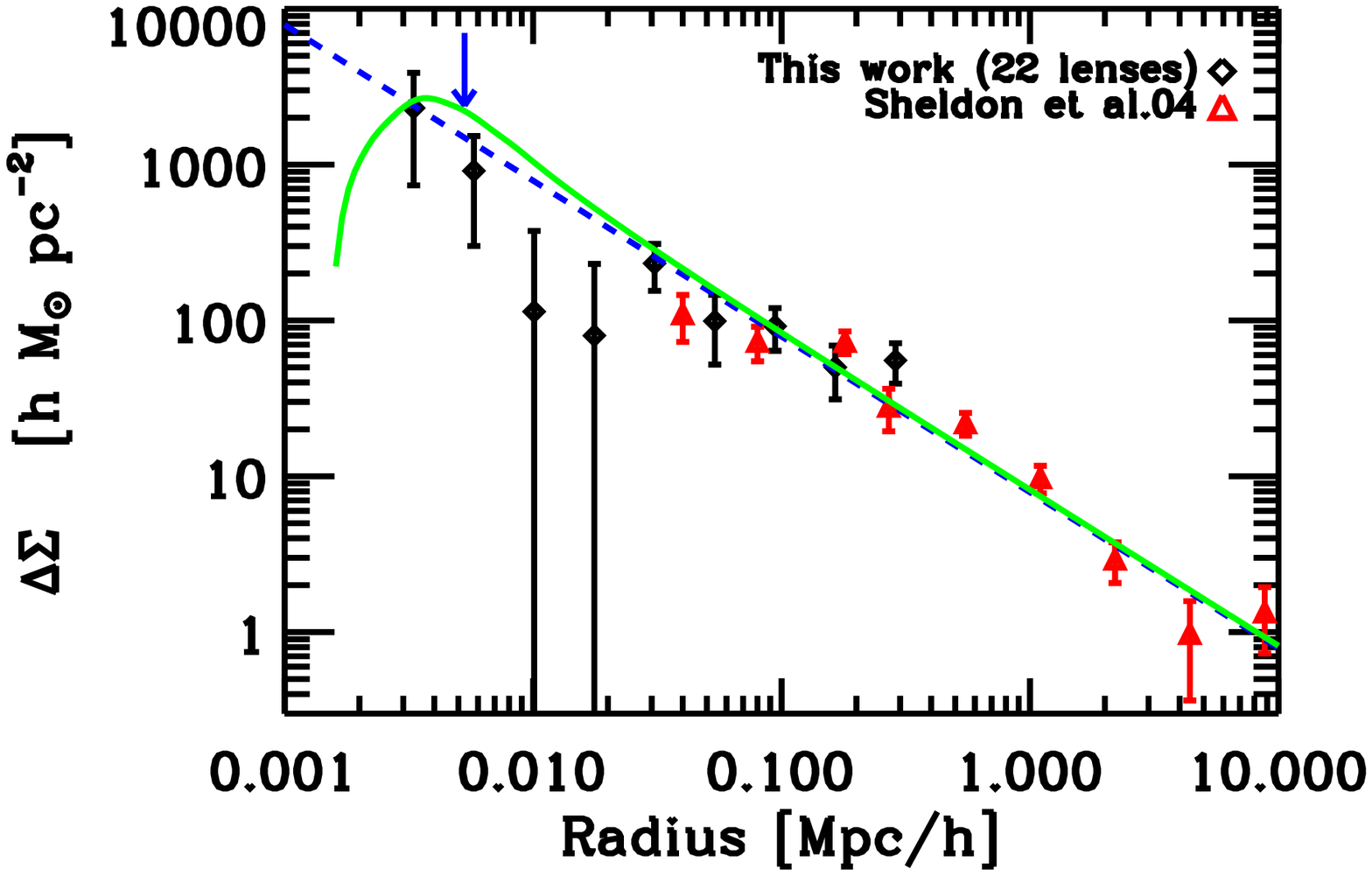}
   \caption{\small Radial shear profile around 22 SLACS strong lenses
  (black diamonds) expressed as the excess surface mass density
  $\dsig(R)$. SLACS bin are completely uncorrelated.  The blue arrow
  denotes the mean effective radius of the lenses stellar component
  $\overline{R_{\rm eff}}\simeq 5.3\hmkpc$. The average critical
  density is $\scrit \simeq 4620\,h\,\msun\,\pc^{-2}$. Previous
  results from S04 scaled to the same average square
  velocity dispersion are shown for comparison as red triangles (see
  \S~\ref{sec:prof-maps} for details). The curves show the shear
  profile expected from strong lensing SIS models. {\it This is not a
  fit}. The solid green curve takes into account non linear effects in
  the relation $\langle e\rangle = g = w \gamma/(1 - w\kappa)$ whereas
  the dashed blue line neglects non linearities. This shows that
  $\langle e \rangle = \overline{w} \gamma$ is a good approximation
  for our purposes.}
  \label{fig:shearprof}
\end{figure}

Measured $\dsig$ values around SLACS lenses are reported in
Tab. \ref{tab:dsigvalues} and shown in Fig~\ref{fig:shearprof}.
The detection significance is derived as follows.
The $\chi^2$ of the data with respect to the
zero shear hypothesis is 47.8 for 9 degrees of freedom. The
probability of finding a higher $\chi^2$ is $3\times10^{-7}$, thus the
non-detection hypothesis is rejected at the 99.99997\% level. For a
Gaussian distribution this is equivalent to 5$\sigma$.

To compare with previous studies, we consider the measurement from S04
for their subsample of massive $\sigma_v>186\kms$. The mean square
velocity of their sample is $\langle\sigma_v^2\rangle^{1/2}\simeq
225\kms$. In order to compare with our points we need to correct for
the different velocity dispersion. Assuming an isothermal profile, the
shear scales as the velocity dispersion squared, so that we need to
scale their points up by $(248/225)^2\sim 1.21$ for a proper
comparison. After this correction, the agreement is excellent in the radial
range $60\lesssim R \lesssim 200 \hmkpc$ of overlap as shown in
Figure~\ref{fig:shearprof}. We also check that our results are in
agreement with the $\dsig$ profile of the {\tt sm7} (early-type)
stellar mass bin of M06.

\begin{table*}[htbp]
  \centering
  \caption{Measured excess surface density $\dsig$ and shear $\gamma$}\label{tab:dsigvalues}
  \begin{tabular}{ccccccc}\hline\hline
    Proj. Radius & $\dsig$ & $\dsig_\times$ & $\sigma_{\dsig}$ & $\gamma_t$& $\gamma_\times$& $\sigma_{\gamma}$\\
    $\hmkpc$  & $\h\msun/\pc^2$ & $\h\msun/\pc^2$ & $\h\msun/\pc^2$ &&&\\\hline
 3.3 & 2307 & -1015 & 1570 &  0.339 & -0.054 &  0.093 \\
 5.8 & 918 &  -526 &  614  &  0.206 & -0.095 &  0.078 \\
 10.1 & 115 &   141 & 264  &  0.074 &  0.010 &  0.047 \\
 17.6 &  81 &  -180 & 153  &  0.028 & -0.011 &  0.029 \\
 30.8 & 232 &   -36 & 80   &  0.064 & -0.005 &  0.017 \\
 53.9 & 100 &    36 & 46   &  0.019 &  0.001 &  0.010 \\
 94.1 &  90 &   -45 & 27   &  0.016 & -0.010 &  0.006 \\
 164.5 & 52 &    52 & 17   &  0.010 &  0.014 &  0.005 \\
 287.5 & 60 &    34 & 17   &  0.014 &  0.008 &  0.004 \\\hline
  \end{tabular}\\
The curl component $\dsig_\times= \scrit \gamma_\times$ is statistically consistent with zero. $\sigma_{\dsig}$ is $1\sigma$ error on both $\dsig$ and $\dsig_\times$. $\sigma_{\gamma}$ is $1\sigma$ error on both $\gamma_t$ and $\gamma_\times$.
\end{table*}

\section{Joint Strong \& Weak lensing modeling}
\label{sec:modeling}

In this section we take advantage of the availability of both strong
and weak lensing constraints to investigate the mass profile of SLACS
lenses from a fraction of the effective radius to 100 effective radii ($3 \lesssim R\lesssim300 \hmkpc$).

\subsection{SIS consistency check}
\label{sec:ccheck}

Before considering more sophisticated models for the density profile,
we first check if the singular isothermal density profile favored in
the inner parts of galaxies by strong lensing alone
\citep[\eg][]{rusin03}, and by strong lensing and stellar dynamics
\citep[paper III]{treu04}, is consistent with our weak lensing
measurements. 

\subsubsection{Consistency with strong lensing}

For a singular isothermal sphere (SIS) the convergence profile as a
function of radius $R$ is:
\begin{equation}
  \kappa(R) = \frac{\rein}{ 2 R} = \gamma(R)\:,
\end{equation}
with $\rein = 4 \pi (\sigma_{\rm SIS}/c)^2 \dlsdos$ in radians and
$\sigma_{\rm SIS}$ the lensing-inferred velocity dispersion which turns out
to be very close to the stellar velocity dispersion $\sigma_*$ of the lens
galaxy (paper II). For a proper comparison, weak and strong lensing
measurements have to be renormalized to the same source plane. Hence
the Einstein radius given by strong lens modeling has to be rescaled
by a factor $\overline{w}_{\rm WL}/w_{\rm SL}$ (see Table
\ref{tab:sample}). Fig. \ref{fig:shearprof} shows that after this
scaling, but {\it without fitting any free parameter}, the strong
lensing SIS models provide a reasonably good description of SLACS weak
lensing data out to  $\sim100 \hmkpc$
(with $\chi^2/{\rm dof}\simeq22.8/9\simeq2.5$), and of the SDSS data
beyond that (with $\chi^2/{\rm dof}\simeq 27.6/9\simeq 2.7$).
Two models are shown: one that neglects the non-linear relation between
reduced shear and
ellipticity (blue dashed) and one that takes this effect into account
as well as the associated non-linear dependence on the source redshift
distribution (solid green). The two curves differ by less than the
error bars of our measurements, showing that a linear relation between
ellipticity and shear is a reasonable approximation given the present
statistical errors.  This analysis shows that the {\it total} mass
density profile of the SLACS lenses is very close to an isothermal
sphere with velocity dispersion equal to the stellar velocity
dispersion. Since the luminous component is steeper than isothermal
outside the effective radius, this finding implies the presence of an
extended dark matter halo which is in turn shallower than isothermal
at similar radii.

\subsubsection{Consistency with strong lensing and stellar dynamics}

\begin{figure}[htb]
  \centering \includegraphics[width=8.6cm]{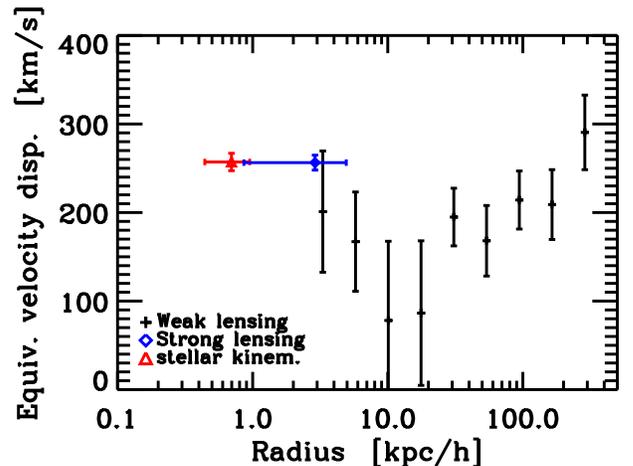}
  \caption{\small Illustration of the complementarity of three mass
  tracers (stellar velocity dispersion, effective strong and weak
  lensing velocity dispersion) over almost three decades in
  radius. Stellar velocity dispersion constrains the innermost regions
  ($\ll \rein$), strong lensing the region around the Einstein
  radius, while weak lensing is most effective beyond the Einstein
  radius.}
  \label{fig:idefix}
\end{figure}

A simple -- although model dependent -- way to compare on the same
plot the mass measurement obtained with strong lensing, stellar
dynamics and weak lensing is obtained in the following manner. For
each radial bin we can define an effective weak lensing velocity
dispersion as the velocity dispersion of the singular isothermal
sphere that reproduces the shear in that bin. The effective weak
lensing velocity dispersion profile is shown in Figure\ref{fig:idefix}
together with the average stellar velocity dispersion determined from
SDSS spectroscopy and the average stellar velocity dispersion of the
singular isothermal ellipsoid that best fits the strong lensing
configuration.

The figure illustrates the complementarity of the three mass tracers,
stellar velocity dispersion well inside the Einstein Radius, strong
lensing at the Einstein radius, and weak lensing outside the Einstein
radius, as well as the dynamic range of the measurement, almost three
decades in radius. The very close correspondence of the stellar and
strong lensing measurement was discussed in paper II, and is confirmed
here for a larger sample of lenses. This paper shows that, albeit with
larger uncertainties, the weak lensing data show that the profile is
approximately flat for another two decades in radius. This is a
qualitative statement as a full joint (strong+weak) lensing and
dynamical analysis is needed to combine the three diagnostics
properly. The three-pronged analysis is beyond the scope of this paper
and is left for future work. In the rest of this paper we will focus
on combining strong and weak lensing in the context of a two-component
mass model.

\subsection{Two component models}
\label{sec:SWLtech}

In the rest of this section we perform a joint strong and weak lensing
analysis of the data, in order to disentangle the mass profile of the
stellar component and of the surrounding dark matter halo. For this
purpose we adopt a simple two component model as detailed below. For
simplicity, we assume that all lenses are at the center of their halo
and none of them is an off-center satellite in a bigger halo. This
approximation is well motivated by the galaxy-galaxy lensing results
of M06 who found that only a small fraction ($\lesssim 15\%$) of
massive ellipticals do not reside at the center of their host halo.
Strictly speaking, the quantity measured by weak lensing is
the projected galaxy-mass cross-correlation function rather than the
actual shear profile of an individual halo. However, the
interpretation of this cross-correlation function within the
successful framework of the ``halo model'' \citep[\eg][]{cooray06}
allows one to disentangle the contribution of the proper halo attached
to a given galaxy (1-halo central term), the halo of a more massive
host galaxy (or group or cluster) if this galaxy is a satellite
(1-halo satellite term), and the contribution due to clustering of
neighboring halos about the main halo attached to the galaxy (2-halo
term).  However, the 2-halo terms only provide a significant
contribution to the galaxy-mass cross-correlation function beyond a
few Mpc (as compared to the outermost $\sim300 \hmkpc$ radial bin
probed here) and our lenses are massive ellipticals and thus most
likely central galaxies. Therefore for the purpose of this analysis,
and given the measured uncertainties, we can assume with little loss of
accuracy that the measured shear profile is representative of the only
surrounding main halo
\citep[see \eg Fig.~1 of][]{mandelbaum05}.

\subsubsection{Model definition}

We model the stellar component with a de Vaucouleurs density profile
\citep{devauc48,maoz93,keeton01soft2} The effective radius of the
stellar component is fixed to the ACS surface photometry. Thus the
only free parameter needed to measure the luminous component is the
average stellar mass-to-light ratio $\Upsilon_V$. The dark matter halo
is assumed to be of the NFW form \citep{NFW97,bartelmann96,wright00}
in which the density reads
\begin{equation}
  \rho(r) = \rho_c \delta_c \left(\frac{r}{r_s}\right)^{-1} \left( 1+\frac{r}{r_s}\right)^{-2}\,.
\end{equation}
with $r_s$ the scale radius and $\rho_c$ the Universe critical
density. The term $\delta_c=\frac{\Delta}{3}c^3 /
\left[\ln(1+c)-c/(1+c)\right]$ relates the so-called concentration
parameter $c=r_\Delta/r_s$. We assume an overdensity of $\Delta=119$
so that $r_\Delta$ can be considered the ``virial'' radius \citep{bryan98}.
This definition agrees with those of \citet{hoekstra05a} and
\citet{heymans06} for our assumed $\Lambda$CDM cosmology. Because
statistical errors are still large with only 22 SLACS lenses used here,
we lack the sensitivity to constrain the dark matter profile in detail.
Thus, instead of fitting a free concentration parameter, we assume the
general relation observed in numerical simulations:
\begin{equation}
  c= \frac{9}{1+z} \left( \frac{M_{\rm vir}}{8.12 \times 10^{12} \hmmsun}\right)^{-0.14} \,,
\end{equation}
\citep{bullock01,eke01,hoekstra05a}. In addition, we are not able to
constrain the virial mass of each lens individually, so we need to
assume a scaling relation between virial mass and V band luminosity of
the form: $M_{\rm vir} = \tau_V L_V $. Note that we check that
assuming a steeper relation $M_{\rm vir} \propto L_V ^{1.5}$
\citep{guzik02,hoekstra05a,mandelbaum06} yields comparable results
because the SLACS lenses span a narrow range in luminosities, with 0.2 dex
rms around $\langle L_V\rangle=5.70\times 10^{10} \hmlsun$.  In
conclusion, our model has only two free parameters, the mass-to-light
ratio of the luminous component $\Upsilon_V$ and the virial
mass-to-light rato $\tau_V$.

Having defined the model, we now define the merit function that will
be used to determine the best fitting parameters with their
uncertainties.  Detailed strong lensing analysis of multiply imaged
sources have shown that extremely tight constraints can be set on the
Einstein radius of individual lenses (\eg paper III), with typically a
few percents relative accuracy $\sigma_{\rein}/\rein\simeq 5\% $. This
can be interpreted as a surface mass measurement since the mean
density $\Sigmabar$ within this radius is by definition equal to the
critical density $\scrit$. Therefore for each lens we are able to
write:
\begin{equation}\label{eq:scrit-rein}
  \Sigmabar(<\rein) = \scrit(\zl,\zs) \;.
\end{equation}
The relative error on $\rein$ translates into
$\sigma_{\Sigmabar}/\Sigmabar\simeq 5\%$. We can thus define a strong
lensing merit function:

\begin{equation}\label{eq:chi2sl}
  \chi^2_{\rm sl} = \sum_{i=1}^{N_{\rm lens}} \left( \frac{{\scrit}_{,i}
  - \Sigmabar_{*,i}(\Upsilon_V) - \Sigmabar_{DM,i}(\tau_V)
  }{\sigma_{\Sigmabar}}\right)^2
\end{equation}
where subscripts $*$ and $DM$ stand for luminous and dark matter
components, evaluated at position ${\rein}_i$.

In a complete analogy, we define the weak lensing  merit function:
\begin{equation}\label{eq:chi2wl}
  \begin{split}
    \chi^2_{\rm wl}  &= 
    \sum_{j=1}^{N_{\rm rbin}} \frac{1}{\sigma_{\dsig,j}^2} \\ 
    & \left[\dsig_j -  \frac{1}{N_{\rm lens}}\sum_{i=1}^{N_{\rm lens}} \left( \dsig_{*,ij}(\Upsilon_V) + \dsig_{DM,ij}(\tau_V) \right) \right]^2
  \end{split}
\end{equation}

where $N_{\rm rbin}$ is the number of radial bins $r_j$ at which
$\dsig_j=\dsig(r_j)$ is obtained from the weak lensing data shown in
Fig. \ref{fig:shearprof}. 

In the next section we will derive the best fitting $\{\Upsilon_V,\tau_V\}$ values that minimize the total $\chi^2=\chi^2_{\rm sl} + \chi^2_{\rm wl}$.

\subsubsection{Results}
\label{sec:NFWres}

\begin{figure}[htb]
  \centering
  \includegraphics[width=8.6cm]{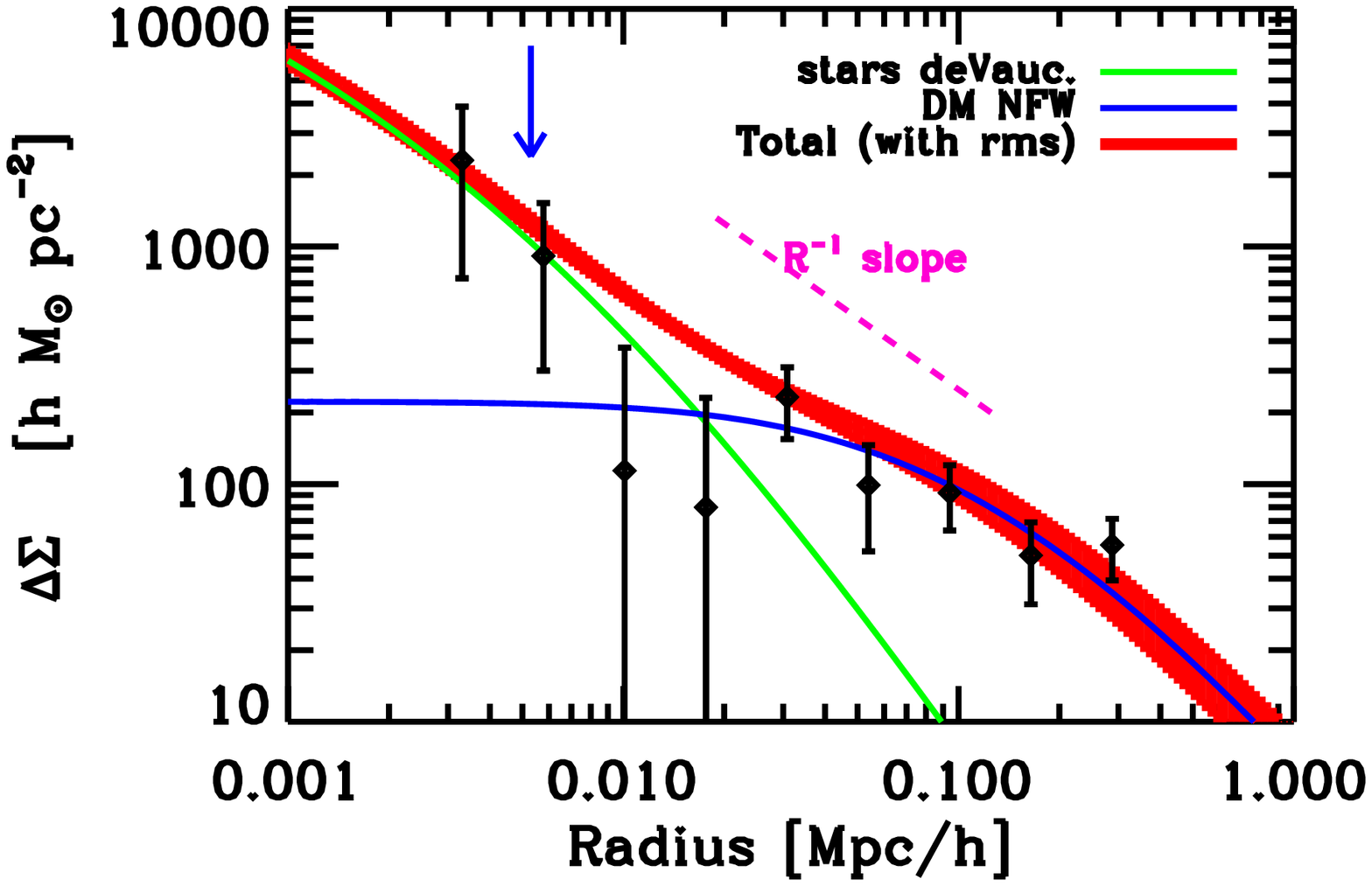}
  \includegraphics[width=8.6cm]{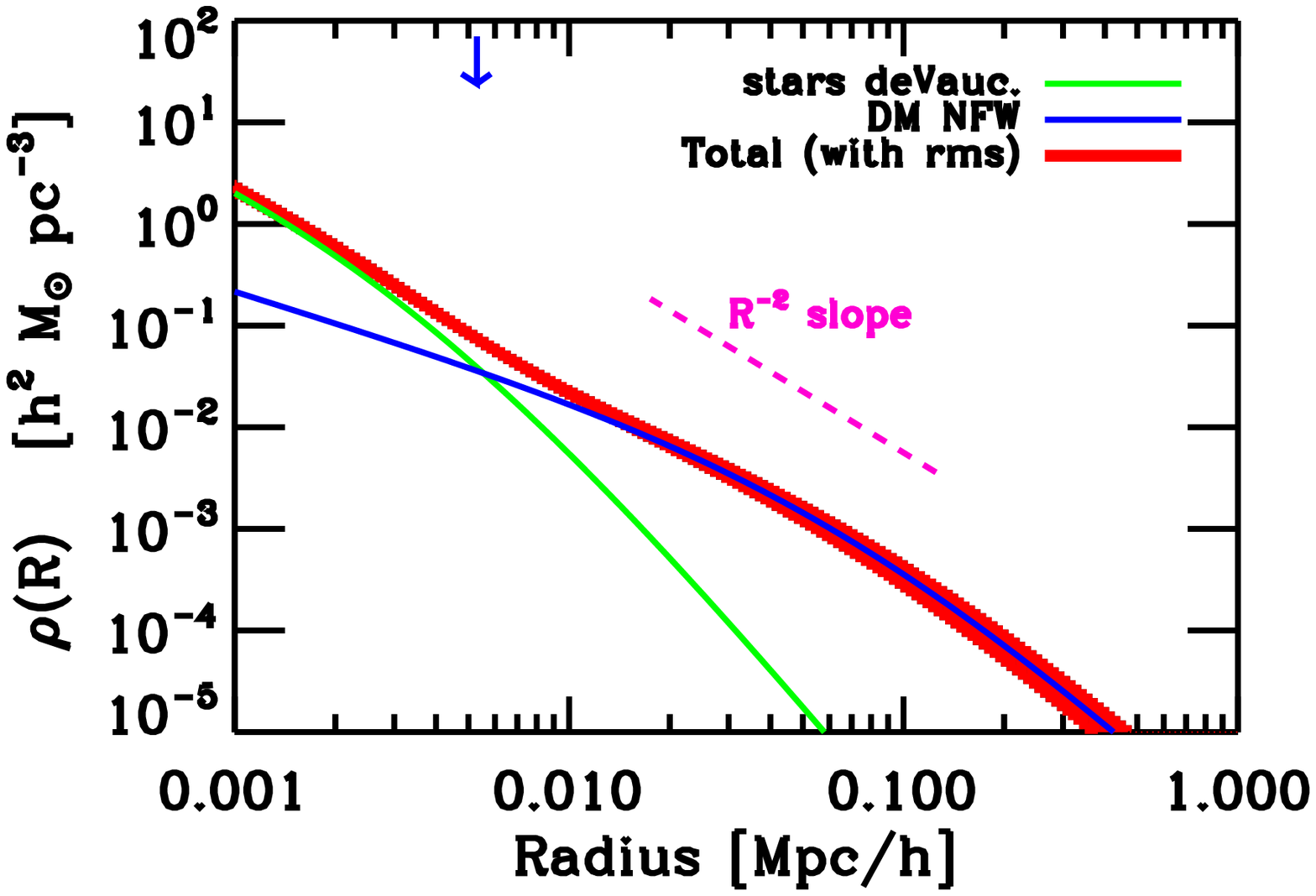}
  \caption{\small {\it Upper panel:} Shear profile (\ie $\dsig$) for the best DM + de Vaucouleurs profile. The contribution of each mass component is detailed (green and blue for stars and DM respectively). The thickness of the red curve codes for the $1\sigma$ uncertainty around the total shear profile. Uncertainties are very small below 10 kpc because of strong lensing data that cannot be shown here. {\it Lower panel:} Similar coding for the three-dimensional density profile $\rho(R)$. The transition between star- and DM-dominated mass profile occurs close to the mean effective radius (blue arrow). The total density profile is close to isothermal over $\sim 2$ decades in radius.}
  \label{fig:best-nfw0}
\end{figure}

The upper panel of Fig. \ref{fig:best-nfw0} shows the radial profile of the
shear for the best fit model together with weak lensing data
points. The fit is excellent with a $\chi^2/{\rm
dof}=29.1/31\simeq0.94$. We see the detail of the contribution of
stellar and dark matter components. This joint strong+weak lensing
analysis allows to disentangle the contribution of each. Because the
latter component is less concentrated and more extended than the
former, there is a radial range $R\sim 20 \hmkpc$ at which surface
mass density flattens. This implies a fast drop in the shear profile
$\dsig(R)$ at that scale which is quite easy to detect.

In the lower panel of Fig. \ref{fig:best-nfw0} we show the
corresponding three-dimensional mass density profile $\rho(R)$. This
profile is close to isothermal although it is made of two components
which are not isothermal. The components combine to make an
almost isothermal density profile at scales $R\lesssim 100\hmkpc$ with
a transition from star-dominated to dark matter dominated profiles
occurring close to the effective radius.

\begin{figure}[htb]
  \centering
  \includegraphics[width=8.6cm]{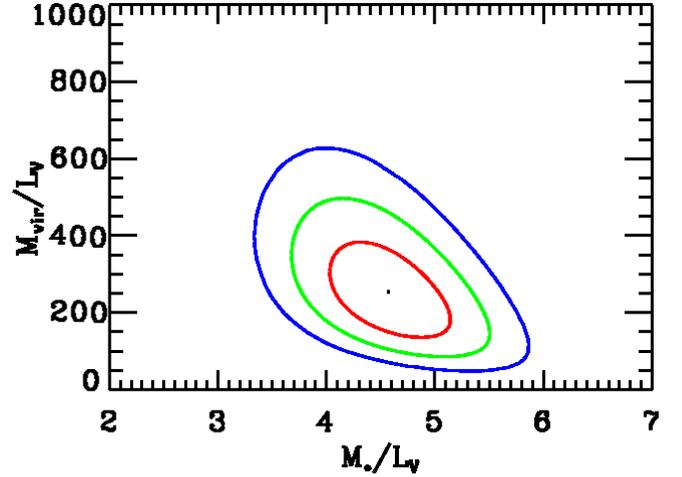}
  \caption{\small Confidence levels around model parameters (1, 2 and
  3 $\sigma$ contours) for the relation between total virial
  mass-to-light ratio $\tau_V=M_{\rm vir}/L_V$ and stellar
  mass-to-light $\Upsilon_V=M_*/L_V$ ratio. Given the sample mean
  luminosity $\langle L_V\rangle =5.7\times 10^{10} \h^{-2}L_\odot$,
  we find a mean sample stellar mass $\langle M_* \rangle =
  2.55\pm0.27 \times 10^{11}\hmmsun $ and virial mass $\langle M_{\rm
  vir}\rangle=14\mypm{6}{5}\times 10^{12} \hmmsun$.}
  \label{fig:cont-nfw0}
\end{figure}

The best fit NFW + de Vaucouleurs lens model yields a stellar
mass-to-light ratio $\Upsilon_V=4.48\pm 0.46 \h\msun/L_\odot$
and virial mass-to-light ratio $\tau_V=246\mypm{101}{87}
\h\msun/L_\odot$. This translates into a virial to stellar mass ratio
$M_{\rm vir}/M_* = 54 \mypm{28}{21}$. Note that stellar mass-to-light
ratio depends in the same way on $h$ as virial mass as they are
inferred from lens modeling and not from stellar evolution
models. Thus $M_{\rm vir}/M_*$ is independent of $h$.
Fig. \ref{fig:cont-nfw0} shows the 68.3, 95.4 and 99.7\%
confidence levels contours for the best fit model parameters. 
A model having a constant mass-to-light ratio (\ie $M_{\rm vir}/L_V=0$)
is ruled out at $\sim 4 \sigma$.

Given the sample mean luminosity $\langle L_V\rangle =5.7\times
10^{10} \h^{-2}L_\odot$, we find a mean sample stellar mass $\langle
M_* \rangle = 2.55\pm0.26 \times 10^{11}\hmmsun $ and virial mass
$\langle M_{\rm vir}\rangle=14\mypm{6}{5}\times 10^{12} \hmmsun$. This
translates into a mean virial (resp. scale) radius
$r_{\rm vir}=393\mypm{47}{53} \hmkpc$ (resp. $r_s=58 \pm 8\hmkpc$).
We note that the virial radius is typically larger that our field of
view, and therefore virial masses rely on extrapolations of our
results. Therefore, we also present more robust measurements like the
projected and three-dimensional mass within a reference radius
$R=200\hmkpc$. Lens modeling yields $M_{3D}(<200\hmkpc) =
(8.1\pm1.8)\times 10^{12} \hmmsun$ and a projected mass
$M_{2D}(<200\hmkpc) = (10.8\pm2.7)\times 10^{12} \hmmsun$ (68\% CL
errors).

To compare with local measurements we convert our $\Upsilon_V$
mass-to-light ratio to the rest-frame B band.
Assuming a typical $(B-V)=0.96$ color for
Ellipticals \citep{fukugita95}, $(B-V)_\odot =0.65$ and a Hubble
constant $h=0.7$, one finds $\Upsilon_B = 4.17 \pm
0.44\,\msun/\lsun$. Using paper II, \citet{treu04} and similar
findings \citep{treu02,treu05a,wel05,diserego05} for the passive evolution of massive early-type
galaxies, we get a redshift zero B band stellar mass-to-light ratio
$\sim 5.81 \pm 0.61$ which is statistically consistent with local
estimates such as $\Upsilon_B = 7.8 \pm 2.7 $ and $\Upsilon_B = 7.1 \pm
2.8$ from \citet{gerhard01} and \citet{trujillo04} respectively. The
low value of $M_*/L$ is also in broad agreement with stellar evolution
models of \citet{bruzual03}, although detailed comparisons depend on
the assumed Initial Mass Function (IMF).

Our modeling can be used relate the V band luminosity within the
Einstein radius $\overline{L}_V(<\rein)$ and the fraction of dark
matter in the same projected radius $f_{DM,2d}(<\rein)$. Using
Eq. \eqref{eq:scrit-rein}, for a given stellar mass-to-light ratio
each lens must verify
\begin{equation}\label{eq:fdm-rein}
  f_{\rm DM,2d}(<\rein) = 1 - \Upsilon_V \overline{L}_V{(<\rein)} / \scrit \,.
\end{equation}

\begin{figure}[htb]
  \centering
  \includegraphics[width=8.6cm,height=5cm]{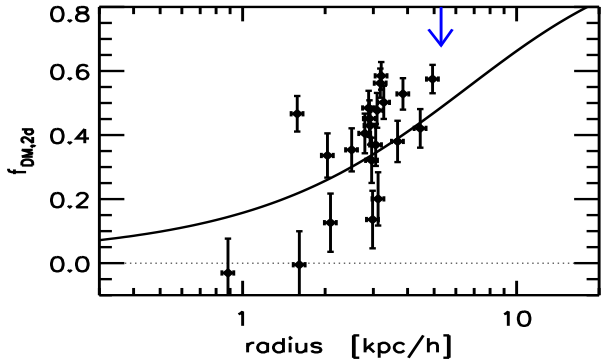}
  \caption{\small Projected dark matter fraction in SLACS lenses as measured from strong lensing at the Einstein radius (data points) and our constraints on the stellar $M_*/L$ ratio. The Einstein radius is expressed in physical units and the solid curve shows the best fit profile of the dark matter fraction $f_{\rm DM,2d}$ inferred from the parametric NFW + de Vaucouleurs modeling of strong+weak lensing data. The mean effective radius is shown as a blue arrow. We see that $f_{\rm DM,2d}(<\reff)\simeq 50\%$ of the projected enclosed mass is in the form of dark matter within the effective radius.}
  \label{fig:fdm}
\end{figure}

Fig. \ref{fig:fdm} shows the inferred projected $f_{\rm DM,2d}$ using
our best fit $\Upsilon_V=4.48\hmmsun/\lsun$. The mean dark matter
fraction within the Einstein radius $\langle f_{\rm
dm,2d}\rangle=0.37\pm0.04$ with $18\%$ rms scatter. Extrapolating to
the effective radius, about half of the projected mass is in the form
of dark matter. The result from the NFW + de Vaucouleurs parameterization
is shown as the solid curve which matches the data points well (see
also papers II and III). This parameterization also allows the deprojected
 dark matter fraction to be calculated, and is found to be $\sim 27\%$
(dotted blue line) within $\reff$. The local deprojected DM and
stellar densities are of the same order at that radius.

It may seem that the two data points with $f_{\rm dm,2d}\simeq 0$ are
responsible for our inferred low $M_*/L$. Since these two lenses have
the most elongated stellar component\footnote{$q_*=(b/a)_* \sim 0.5$
whereas the other lenses have a mean $q_*=0.81$ and dispersion 0.08},
the assumption of circular symmetry may break down for
them. Furthermore, if they have a disk component with younger starsq,
this could reduce their global $M_*/L$ with respect to that of pure
spheroidal systems. However, redoing the strong+weak lens modeling
without these does not change the stellar mass-to-light ratio to
significantly ($M_*/L_V=4.90\pm0.53 \hmmsun/\lsun$).

\subsection{Comparison to previous galaxy-galaxy lensing works}
\label{sec:NFW-compar}

We first compare our findings to the SDSS weak galaxy-galaxy lensing
analysis of M06 who defined $M_{\rm cent}$ as the mass enclosed in a
sphere within which the mean density is 180 times the background
density, similar to our definition. Our lens sample should be compared
to the {\tt sm7} stellar mass bin for early-type galaxies with
$\langle M_*\rangle =39.6\times 10^{10} \msun $. It would also lie
between {\tt l6b} and {\tt l6f} early-type galaxy luminosity bins with
a typical $\langle L\rangle \simeq 6.3 L_*$. To convert the rest frame V
band to the SDSS $r'$ band we use $V-r'\simeq 0.36$ for Ellipticals
\citep{fukugita95}. Since at the exponential tail of the luminosity
and mass functions one is highly sensitive to the scatter in the
mass-luminosity relation, M06 applied corrections calibrated into
simulations \citep{mandelbaum05} whereas our analysis does not attempt
to correct for this effect. For the {\tt l6f} bin, $ M_{\rm cent}/L_r
= 674 \mypm{210}{203} \h \msun/\lsun$ as found by M06 yields $M_{\rm
cent}/L_V = 896 \pm{275} \h \msun/\lsun$ (95\% CL). Likewise the {\tt
sm7} bin of M06 yields $M_{\rm cent}/M_*=256\mypm{44}{68}$ (95\% CL
and using $h=0.7$). These values are significantly larger than our
results. However if we apply a similar correction as these authors our
virial mass should be raised by 66\%, yielding $M_{\rm
vir}/L_V=408\mypm{168}{144}\h \msun/\lsun $ and $M_{\rm
vir}/M_*=89\mypm{46}{35}$. This latter correction thus brings our
findings into statistical agreement. The small difference might
be due to our inability to probe the very outer parts of halos and
efficiently constrain virial masses. However, we emphasize that the
availability of strong lensing constraints put tight constraints on
the column density enclosed by $\rein$, which means that the outer
parts of halos cannot contribute much in making lens galaxies
critical, or perhaps to the fact that we fit for $M_*/L$, while M06
use the value determined from stellar population synthesis
models. Issues related to comparisons between \citet{guzik02} and M06
results are addressed in the latter paper. In any case, we find a
much better agreement between our $M_{\rm vir}$ measurement and those
of \citet{guzik02} which give $M_{\rm vir}/L_V \simeq 296\pm 51
\h\msun/\lsun$ after matching our lens sample selection.

We now compare our analysis with the \citet{hoekstra05a} results. Since we
chose to match their definition of virial mass and concentration, we
expect comparisons to be easier although the mean redshift of their
lens sample is $\sim 0.32$. Our lens sample would passively brighten
to $\langle L_V\rangle\simeq 6.3\times 10^{10} \hmlsun$ at $z=0.32$
which is $\sim 2.45$ times brighter than their higher luminosity bin
having $L_V\simeq 2.45\times 10^{10} \hmlsun$. Therefore we need to
extrapolate their results using their $M_{\rm vir} \propto L^{1.5}$
relation. They find $M_{\rm vir}/L_V\simeq 253\mypm{38}{35}\h \msun/\lsun$.
This value is statistically consistent with ours. However, a detailed
comparison is made difficult due to the fact that
the authors mix early- and late-type galaxies and they avoid lens galaxies
in dense environments. Using stellar evolution models, they estimate the
virial to stellar mass in their reddest subsample $(B-V)_{\rm
rest}\sim 0.95$ to be $M_{\rm vir}/M_* \simeq 43 \pm 6$ for a scaled
Salpeter IMF \citep{bell01} or $M_{\rm vir}/M_* \simeq 27 \pm 4$ for a
PEGASE Salpeter IMF. The Scaled Salpeter IMF hypothesis turns out to be in
better agreement with our measurements. In addition it predicts stellar
mass-to-light ratios $\Upsilon_B\sim 4.5$ (Solar units) closer to our estimates
than the PEGASE IMF which predicts $\Upsilon_B\sim 6.5$.

Finally, \citet{heymans06} measure virial masses of lens galaxies in
the range $0.2\le \zl \le 0.8$ in a narrow range of luminosity $L_r =
2.4 \times 10^{10} \hmlsun $ with a 0.2 dex dispersion about this
mean. This sample is dominated by early-type galaxies.  Again,
extrapolation to our sample mean luminosity is somewhat uncertain but,
using the $M_{\rm vir} \propto L_V^{1.5}$ scaling, their results give
$M_{\rm vir}/M_* = 76\pm 25$ which is in excellent agreement with our
$M_{\rm vir}/M_* = 54 \mypm{28}{21}$.

These comparisons show that our virial mass estimates are in good
agreement with other studies after extrapolation of our constraints on
the radial shear profile (ending around $\sim300\hmkpc$) out to the virial
radius $\sim 480\hmkpc$. Comparing with other results obtained for
less massive systems on average increases uncertainties. However our
results on the halo virial masses are well consistent with this
ensemble of results above as they lie in between them. In addition,
the measured shear profile remarkably match those of S04 and M06 in
the radial range $30\lesssim R\lesssim300\hmkpc$. This gives a
valuable support to the validity of our results and the control of
systematic effects.

\section{Discussion}
\label{sec:discuss}

Our joint weak+strong lensing modeling of SLACS lenses with a two-component
mass model allowed us to successfully disentangle the
contribution of each giving sensible results for both the stellar
mass-to-light ratio $M_*/L_V=4.48\pm0.46\h\msun/\lsun$ and virial
mass-to-light ratio $M_{\rm vir}/L_V=246\mypm{101}{87}\h\msun/\lsun$,
in good agreement with other studies. Assuming a NFW and de
Vaucouleurs forms for each density profiles provides a good
description of the data ($\chi^2\simeq0.94$ per degree of freedom).

This analysis shows that the total density profile is close to
isothermal out to $\sim100\hmkpc$. It is now a well established 
from SLACS (paper III) and earlier strong lensing studies
\citep[\eg][]{treu02,rusin03,rusin05b,treu04} that the {\it total}
mass profile of lens galaxies is close to isothermal ($\rho\propto
r^{-2}$) within $\reff$. In paper III this result was established by
combining strong lensing and stellar kinematics. The present analysis
extends and strengthens this results as we find that the dark
matter and stellar components combine themselves to form an isothermal
total density profile well beyond the effective radius.

We find that the transition from a star-dominated to a dark-matter-dominated
density profile must occur close to $\reff$. This peculiar
transition is also observed by \citet{treu04} which combined strong
lensing and stellar kinematics in higher redshift lenses. Similar
results can be found in \citet{mamon05a,mamon05b}. In addition, the
{\it ``mean''} fraction of DM $f_{dm,3D}(<\reff)\sim30\%$ is in
excellent agreement with local estimates
\citep{kronawitter00,gerhard01,borriello03}, and more recently from the
SAURON project \citep{cappellari06}. At this point, it is noteworthy
to note an important result of paper II, that is, strong lensing
galaxies have similar internal properties as normal early-type
galaxies in terms of their location in the Fundamental
Plane. Our findings can thus be generalized.

We emphasize that we did not investigate other parameterizations for
the DM halo. For instance, a steeper profile ($\rho_{DM}\propto
r^{-\alpha}$ with $\alpha>1$) could possibly arise from the adiabatic
contraction \citep{blumenthal86,gnedin04} of a NFW halo which is found
in pure dark matter N-body cosmological simulations made without
taking into account the complex physics of baryons. With the small
sample of lenses we are considering here, we are not sensitive yet 
to that level of detail in the inner slope of the assumed DM profiles. 
We note however that the rather low values of $M_*/L_V$ we find would
make it unlikely to have a dark matter halo much steeper than NFW
($\rho\propto r^{-1}$ at the center) \citep[see
also][]{borriello03,humphray06}. Since it is reasonable to assume that
baryons somehow perturb the DM halo within the effective radius, we
emphasize that our successful NFW parameterization should rather be
considered as a fitting formula for the perturbed halo. In future
papers, with the complete observed lens sample and spatially resolved
measurements of stellar kinematics, we plan to determine with
unprecedented accuracy the inner slope of the DM profile below $\reff$
and at the same time $M_*/L_V$.

The inner regions of lens galaxies can be considered as representative
of the whole parent sample of early-type galaxies as shown in paper II.
Although there is no firm observational evidence, it is thought that
environmental effects may bias the population of lens galaxies since
extra convergence coming from surrounding large scale structure may
boost lens efficiency while leaving the internal dynamics of lens
galaxies unchanged \citep[see also][]{keeton04,fassnacht06a}.
In the present analysis
we address the issue of whether lenses are representative of the
overall population at larger radii, by comparing our weak lensing
results with those obtained for non-lens samples.
The present analysis shows that, on intermediate
scales $\reff< R \lesssim 300\hmkpc$, SLACS massive lenses have the
same shear properties as normal Ellipticals as found by M06 or
S04. We find a good agreement between our virial mass estimates
and semi-analytic predictions like those developed in the ``halo model''
\citep{mandelbaum05,mandelbaum06,cooray06}.
Any systematic environmental effect able to perturb shear
measurement on those scale is below our present observational uncertainties.
When the ACS follow-up is finished, the weak lensing analysis will
provide important new information on the environment of strong lens
galaxies. This issue is deferred for a future work.

In terms of the internal structure of early-type galaxies, the present
analysis strengthens and extends the results presented in paper III
and gives additional support to the picture proposed there for their
formation. The isothermal density profile must be produced at early
stages of their evolution process ($z\gtrsim 2$) by merging/accretion
involving dissipative gas physics to quickly increase the central
phase space density since pure collisionless systems would rather
develop shallower density slopes $\rho \propto r^{[-1, -1.5]}$
\citep[\eg][]{NFW97,moore98,ghigna98,jing00,navarro04}. Once
isothermality is set, early-type galaxies may evolve passively, or
may keep growing quiescently via collisionless ``dry'' mergers or by
accretion of satellites, which preserve the inner density profile of
collisionless materials (stars and/or DM). See discussion in paper III
and references therein for further details.

\section{Summary \& Conclusion}\label{sec:sum}
We demonstrate that, using deep ACS images, it is possible to measure
a weak lensing signal for a modest sample of massive early-type
galaxies ($\sigma_v \sim 250\kms$ or $\log M_*/\msun \sim 11.35 $):
with only 22 lenses we are able to detect shear signal at
5$\sigma$ significance. Key to this success is the large density of well-resolved
background sources afforded by ACS, which beats down shot noise and
reduces the problem of dilution by lens satellites since background
sources greatly outnumber satellites. In addition, these lenses are
very distant from one another and hence completely statistically
independent. Furthermore, special care has been taken to control
systematic errors, using the most advanced techniques to model and
correct for the ACS PSF and other instrumental effects. By analyzing a
sample of 100 blank fields from the COSMOS survey, we show that
residual systematics in the shear measurement are less than 0.3\%.

Although weak lensing alone can provide interesting results on lens
density profiles at intermediate scale, the great power and
originality of this work is the combination of strong and weak lensing
constraints. Modeling weak {\it and} strong lensing in massive
($\langle\sigma_v^2\rangle^{1/2}\simeq 248\kms$) SLACS galaxies as a
sum of a stellar (de Vaucouleurs) plus dark matter halo (NFW)
components, we could disentangle to contribution of the two components
overall mass budget. The main results of this joint analysis can be
summarized as follows:

\begin{itemize}

\item The total density profile is close to isothermal from
$0.5\reff\lesssim R\lesssim 100\reff$ although neither the stellar nor
the dark matter density profile is isothermal.

\item The transition from star- to dark matter-dominated density
occurs at $\sim \reff$, leading to a dark matter fraction within this
radius $f_{dm,3D}(<\reff)=27\pm4 \%$.

\item The best fit stellar mass-to-light ratio is
$M_*/L_V=4.48\pm0.46\h\msun/\lsun$ in agreement with local results and
old stellar populations.

\item The best fit virial mass-to-light ratio is $M_{\rm
vir}/L_V=246\mypm{101}{87}\h\msun/\lsun$ in agreement with
galaxy-galaxy weak lensing results of non-lens galaxies.
We found a mean virial mass and radius
$\langle M_{\rm vir}\rangle=14\mypm{6}{5}\times 10^{12} \hmmsun$
 and $r_{\rm vir}=393\pm 50 \hmkpc$ respectively.

\item The agreement with other weak lensing studies shows that the
outer halos of lenses and non-lenses are consistent within the
errors. In other words, if lens early-type galaxies live in peculiar
environments, their effect on the shear profile down to $300\hmkpc$
from the lens center are below our statistical errors. 
\end{itemize}

We forecast a $\sim 10\sigma$ detection by the end of the ongoing deep
follow-up imaging with {\it HST/ACS}. When completed, the SLACS sample
of lenses with well resolved kinematics will provide valuable constraints
on stellar populations and density profiles of both stellar and dark matter
components down to several hundred kiloparsecs of the lens center,
thus allowing to address internal properties of lens galaxies as well
as the effect of their environment. To complete the picture on
early-type galaxiess and structure formation, it is important to
extend SLACS results to higher redshift by increasing the number
of such strong lenses through new observational efforts.

\acknowledgements 
We thank Konrad Kuijken and Kevin Bundy for useful suggestions
and Phil Marshall for a careful reading of the paper.
We would also like to acknowledge insightful discussions with Alexie
Leauthaud on the redshift distribution of sources in the COSMOS
survey. RG, TT, LVEK, ASB and LAM would like to thank the Kavli
Institute of Theoretical Physics and its staff for the warm
hospitality during the program ``Applications of gravitational
lensing'', when a significant part of the work presented here was
carried out. The work of LAM was carried out at Jet Propulsion
Laboratory, California Institute of Technology under a contract with
NASA. This research is supported by NASA through Hubble Space
Telescope programs SNAP-10174, GO-10494, SNAP-10587, GO-10798,
GO-10886, and in part by the National Science Foundation under Grant
No. PHY99-07949.
TT acknowledges support from the NSF through CAREER award NSF-0642621.
Based on observations made with the NASA/ESA Hubble
Space Telescope, obtained at the Space Telescope Science Institute,
which is operated by the Association of Universities for Research in
Astronomy, Inc., under NASA contract NAS 5-26555.  This project would
not have been feasible without the extensive and accurate database
provided by the Digital Sloan Sky Survey (SDSS).  Funding for the
creation and distribution of the SDSS Archive has been provided by the
Alfred P. Sloan Foundation, the Participating Institutions, the
National Aeronautics and Space Administration, the National Science
Foundation, the U.S. Department of Energy, the Japanese
Monbukagakusho, and the Max Planck Society. The SDSS Web site is
http://www.sdss.org/.  The SDSS is managed by the Astrophysical
Research Consortium (ARC) for the Participating Institutions. The
Participating Institutions are The University of Chicago, Fermilab,
the Institute for Advanced Study, the Japan Participation Group, The
Johns Hopkins University, the Korean Scientist Group, Los Alamos
National Laboratory, the Max-Planck-Institute for Astronomy (MPIA),
the Max-Planck-Institute for Astrophysics (MPA), New Mexico State
University, University of Pittsburgh, University of Portsmouth,
Princeton University, the United States Naval Observatory, and the
University of Washington.

\nocite{conroy07}
\bibliographystyle{aa}
\bibliography{references}

\end{document}